\documentclass[preprint,12pt]{elsarticle}

\usepackage{amssymb}
\usepackage{amsmath}
\usepackage{multirow} 
\usepackage[dvipsnames]{xcolor}
\usepackage{underscore}
\usepackage{longtable}
\usepackage{booktabs}
\usepackage{subcaption}

\journal{Computational Materials Science}

\begin{document}

\begin{frontmatter}

\title{Indirect prediction and screening of frustrated magnetic materials using machine learning}

\author{Regina Burganova$^{a*}$,  Irina Piyanzina $^{b}$, Ruslan Akhmerov$^{c}$, Alexander Sigov$^{d}$, 
Maxim Fedorov$^{a}$  \\ 
i.piyanzina@gmail.com}

\affiliation{organization={$^a$Institute for Information Transmission Problems (Kharkevich Institute)},%Department and Organization
           addressline={Bolshoy Karetny per. 19, build.1}, 
            city={Moscow},
            postcode={127051}, 
%           state={},
           country={Russia}} 
           
\affiliation{organization={$^b$Center of Semiconductor Devices and Nanotechnology, Computational Materials Science Laboratory, Yerevan State University},    
addressline={1 Alex Manoogian St.}, 
           city={Yerevan},
            postcode={0025}, 
      %     state={},
           country={Republic of Armenia}}

           \affiliation{organization={$^c$Progsol LLC},
            addressline={Noyabrskaya, 52~str.},  
            city={Orenburg},
            postcode={460050}, 
            country={Russia}}

\affiliation{organization={$^d$Department of Nanoelectronics, MIREA—Russian Technological University},%Department and Organization
           addressline={78 Vernadskiy Avenue}, 
            city={Moscow},
            postcode={119454}, 
%           state={},
           country={Russia}}

\begin{abstract}
Magnetic frustration is commonly quantified by the frustration index, calculated from the Curie–Weiss ($\Theta$) and N\'eel ($T_\mathrm{N}$) temperatures. Because both quantities are not simultaneously available for most materials, direct prediction of magnetic frustration index remains challenging. Here, we propose a machine learning workflow for indirect prediction of frustration index and consequent screening of potential frustrated magnetic compounds through independent prediction of $\Theta$ and $T_\mathrm{N}$. In our work, composition-based  models implemented via CatBoost algorithm, trained on experimental data were optimized using chemically informed dataset subdivision. The highest predictive accuracy was achieved for the rare earth-oxide model in predicting $T_\mathrm{N}$ ($R^2=0.83$) and for the rare earth model in predicting $\Theta$ ($R^2=0.72$). Ensemble uncertainty was incorporated through a conservative frustration index to improve the reliability of candidate selection. Screening of approximately $67000$ Materials Project compounds identified $20$ high-confidence frustrated magnetic candidates, including several experimentally known frustrated magnets and previously unexplored oxide systems potential for further investigations.
\end{abstract}

\begin{highlights}
\item ML workflow developed for indirect prediction of magnetic frustration.
\item Independent prediction of N\'eel and Curie–Weiss temperatures improves screening.
\item Chemically informed dataset subdivision enhances model performance.
\item Uncertainty-aware ranking identifies robust frustrated-magnet candidates.
\item Screening of around 67,000 compounds yielded 20 high-confidence candidates.
\end{highlights}

%%Graphical abstract
\begin{graphicalabstract}
\includegraphics[width=\linewidth]{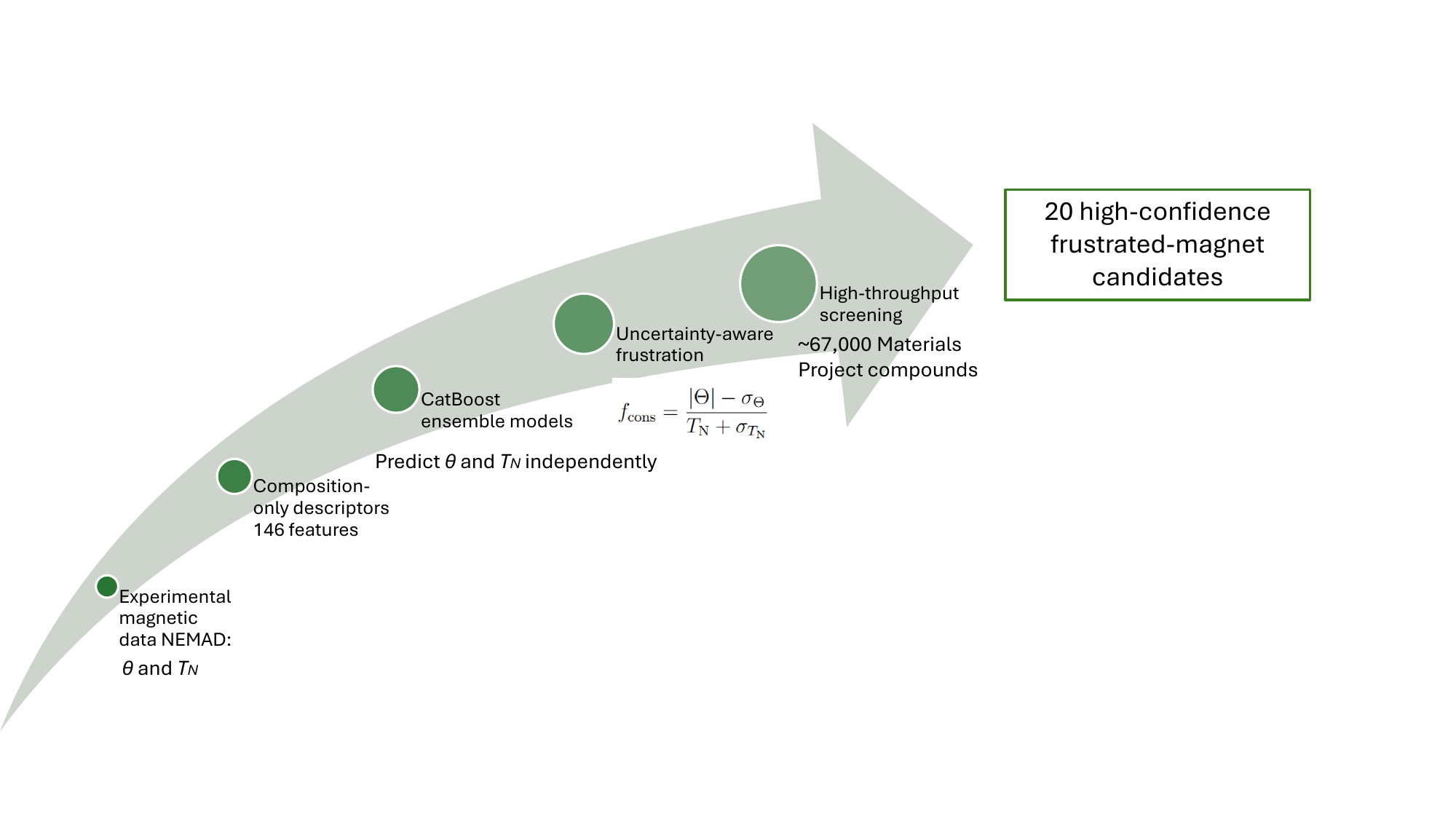}
\end{graphicalabstract}


\end{frontmatter}

\section{Introduction}
\label{intro}
Traditionally, the discovery of magnetic materials has relied on chemically guided synthesis, experimental characterization, and iterative optimization, making the overall process slow and resource-intensive. Determining magnetic ordering temperatures, exchange interactions, and magnetic ground states often requires extensive measurements including magnetic susceptibility, neutron scattering, heat-capacity analysis, and spectroscopic techniques. At the same time, first-principles electronic structure calculations provide microscopic representation of magnetic interactions, but their accurate prediction across broad materials remains computationally demanding and sensitive to methodological approximations, such as the choice of exchange-correlation functional, treatment of electron correlation effects (e.g., Hubbard  corrections), spin-orbit coupling, magnetic configuration sampling, and convergence parameters. Small variations in computational settings may lead to substantial changes in predicted exchange interactions, magnetic anisotropy energies, and magnetic ordering temperatures, especially in strongly correlated and frustrated systems~\cite{desmarais2025meta, garrido2025challenges}. Consequently, rapid identification of promising magnetic compounds remains one of the central challenges in modern materials science.

Recent advances in materials informatics and machine learning (ML) have opened new opportunities for accelerating the discovery of magnetic materials by defining correlations between the composition, structure, and experimentally measured magnetic properties. ML approaches have already been successfully applied for the prediction of magnetic transition temperatures, magnetic response functions, hysteresis behavior, and magnetic ground states using experimentally derived data~\cite{leszczynski2024machine}. Such approaches enable rapid screening of large chemical spaces and significantly reduce the number of compounds requiring detailed experimental investigation. 

Moreover, there are a number of recent studies focused on prediction of magnetic ordering temperatures from experimentally reported magnetic databases. One of the examples is the work of Nelson and Sanvito, who demonstrated that ML models trained on experimentally known ferromagnets (FM) can successfully predict Curie temperatures using composition-based descriptors and random forest regression models~\cite{nelson2019predicting}. Their study established the feasibility of data-driven magnetic screening even without explicit structural information. Subsequently, J.\,F.\,Belot and coauthors presented the prediction of Curie temperature using large experimental datasets of FM compounds and compared several ML algorithms together with different descriptor representations. Their results showed that ensemble learning approaches, particularly, random forest models, provide robust predictive performance while remaining relatively insensitive to highly complicated descriptor sets~\cite{belot2023machine}.

Later studies expanded both the size of the dataset and the complexity of the descriptor space. T.\,Long and coworkers combined the prediction of Curie temperatures with magnetic-ground-state classification of intermetallic compounds using experimentally derived magnetic datasets together with structural and compositional descriptors. Their workflow demonstrated that ML screening can be directly integrated into crystallographic databases for the identification of previously unexplored magnetic materials~\cite{long2021accelerating}. More recently, Jung et al. developed a gradient-boosted feature-selection framework trained on approximately $35\,000$ experimentally reported Curie temperatures collected from literature databases. By combining recursive feature elimination, Bayesian optimization, and composition-based descriptors, the authors achieved predictive accuracies approaching $R^2= 0.92$, emphasizing the importance of the dataset curation and physically meaningful feature engineering for the prediction of magnetic properties~\cite{jung2024machine}.

Another notable example is the physics-informed ML framework proposed by Singh et al., who trained a random forest model on experimentally measured Curie temperatures of rare-earth magnetic compounds and subsequently validated the predictions through targeted synthesis of previously unexplored Ce-Zr-Fe intermetallics~\cite{singh2023physics}. In addition to achieving high predictive accuracy, the authors demonstrated that the ML model successfully reproduces physically meaningful trends associated with the de Gennes scaling and electronic-structure characteristics, illustrating how physically informed descriptors and experimental validation can improve the reliability and interpretability of magnetic-property prediction~\cite{singh2023physics}.

In parallel, graph-based and deep-learning approaches have also been used for magnetic materials modeling. Namely, A.\,A.\,Orang and coauthors demonstrated that graph-neural-network-derived representations combined with experimental datasets significantly improves the accuracy of the Curie temperature prediction with respect to conventional descriptor-only approaches. Their work highlighted the growing role of representation learning and graph-based chemical embeddings in magnetic materials informatics~\cite{orang2026predicting}. Additionally, neural network approaches have also been successfully applied to model nonlinear magnetic response and hysteresis directly from experimentally measured magnetic characteristics~\cite{leszczynski2024machine}. Together, these studies demonstrate the growing role of deep learning in magnetic materials research. Nevertheless, for descriptor-based tabular datasets, tree-based ensemble methods remain highly competitive, and CatBoost has recently been shown to outperform fully connected neural networks while providing excellent predictive accuracy and interpretability~\cite{navratil2025interpretable}.

Despite this rapid progress, the majority of ML studies on magnetic materials remain focused on FM and Curie temperature prediction, whereas antiferromagnets (AFM) and frustration-related phenomena remain comparatively underexplored. Nevertheless, magnetic frustration and competing exchange interactions have long been the subject of extensive theoretical and experimental investigations because of their fundamental importance for magnetic phase behavior and functional magnetic properties~\cite{berzin2012switching, sigov2023behavior, sigov2023spin, sigov2023physical}.

Among magnetic materials, frustrated systems represent a particularly important and challenging class. In particular, in frustrated magnets, lattice geometry or competing exchange interactions suppress conventional long-range magnetic ordering, often stabilizing exotic quantum states including spin liquids, spin ices, and dynamically fluctuating correlated phases~\cite{wiebe2024fantastic}. Such systems attract considerable interest because of their connection to emerging quantum phenomena and potential relevance for future quantum technologies. 

Recent studies have further highlighted the potential of frustrated magnetic materials for cryogenic magnetocaloric applications. In particular, geometrically frustrated Gd-based compounds such as Gd$_2$Ti$_2$O$_7$ have demonstrated notable magnetocaloric properties at low temperatures~\cite{zhang2024geometrically}, while Gd$_2$B$_2$MoO$_9$ has enabled refrigeration down to $0.16$\,K, demonstrating the potential of frustrated magnets for ultra-low-temperature magnetic cooling~\cite{zhang2026refrigeration}. More recently, LiCaGd$_5$(BO$_3$)$_6$ was reported to exhibit a pronounced low-field magnetocaloric response at ultra-low temperatures~\cite{zhang2026geometrically}. Therefore there is an importance of identifying new frustrated magnetic materials with suitable magnetic and thermodynamic properties for cryogenic applications as well.

However, experimental discovery of frustrated materials remains exceptionally difficult due to the enormous diversity of possible crystal structures and magnetic sublattices. Recent topology-based studies demonstrated that only certain classes of magnetic lattices are likely to host strongly frustrated behavior, helping narrowing the search space for candidate materials~\cite{wiebe2024fantastic}.

In experimental studies, magnetic frustration is commonly quantified using the frustration index~\cite{ramirez1994strongly}:
\begin{equation}
\label{eq:f_index}
    f = \frac{|\Theta|} {T_\mathrm{N}},
\end{equation}
where $\Theta$ is the Curie-Weiss temperature and $T_\mathrm{N}$ is the N\'eel temperature. Large values of $f$ indicate strong suppression of long-range magnetic ordering relative to the characteristic magnetic interaction scale and are often considered signatures of frustrated magnetic behavior. Importantly, the two quantities entering this relation probe different physical aspects of magnetism. The Curie-Weiss temperature reflects the overall strength and sign of magnetic exchange interactions, whereas the N\'eel temperature is associated with stabilization of long-range AFM order within a specific crystal and magnetic environment. 

It should be noted that the frustration index is traditionally used to characterize geometrically frustrated magnetic systems. However, in a broader context, it serves as a general measure of competing magnetic interactions. In this work, we adopt this generalized interpretation, allowing the identification of both geometrical and non-geometrical sources of frustration.

As a result, prediction of magnetic frustration is fundamentally more difficult than prediction of a single magnetic observable. Rather than being an independent measurable quantity, the frustration index emerges from the interplay between multiple experimentally determined magnetic energy scales. Consequently, development of ML frameworks capable of simultaneously capturing Curie-Weiss and N\'eel behavior is necessary for data-driven discovery of frustrated materials.

In this work, we propose a ML approach for indirect prediction of magnetic frustration through separate prediction of Curie-Weiss and N\'eel temperatures for AFM compounds. Using experimentally reported magnetic moments together with composition-only-based descriptors, our framework enables high-throughput screening of candidate frustrated materials while remaining directly connected to experimentally measurable magnetic observables. The proposed approach is intended not to replace experimental characterization, but to accelerate it by prioritizing compounds with potentially enhanced frustration for further synthesis and detailed magnetic investigation.

\section{Methods}
\label{methods}
\begin{figure}[h!]
    \centering
   \includegraphics[width=0.9\linewidth]{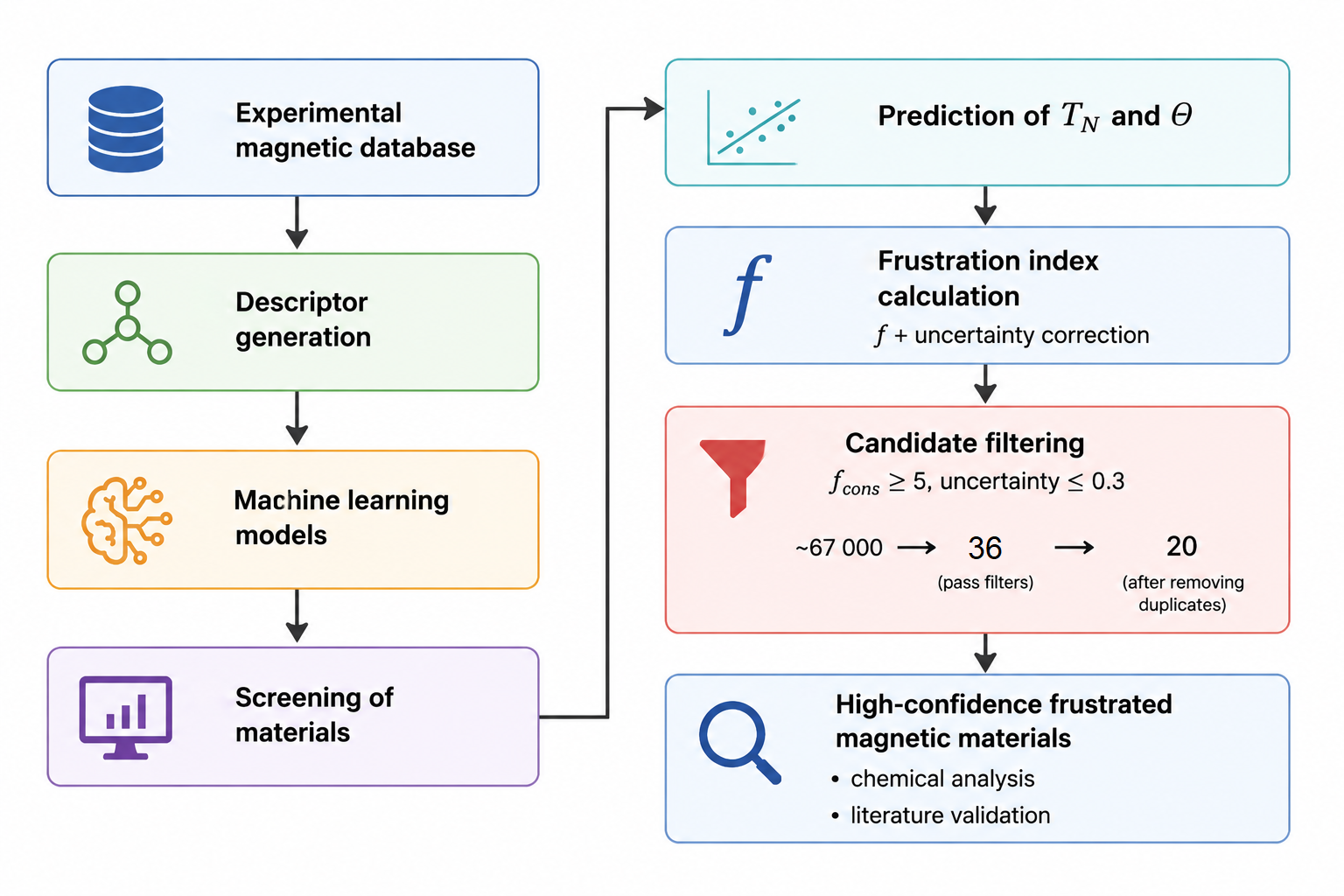}
   \caption{Screening workflow used in the present research.}
   \label{fig:screening_workflow}
\end{figure}

The research workflow is presented in Figure\,\ref{fig:screening_workflow}.
Further, every stage will be described in details. 

\subsection{Dataset processing}
\label{dataset}
The recently introduced North East Materials Database (NEMAD)~\cite{itani2025northeast} was used as the initial source of experimental magnetic data collected from the literature. It provides more than $33\,000$ records of magnetic materials with a total of $19$ features, including compositional, structural, and magnetic characteristics such as magnetic ordering and Curie-Weiss temperatures which we are interested in. 

Two independent regression tasks were formulated: (i) prediction of the N\'eel temperature $T_\mathrm{N}$ and (ii) Curie-Weiss temperature $\Theta$. Since the experimental availability of these values differs significantly across compounds, the datasets for $\Theta$ and $T_\mathrm{N}$ were treated and processed independently. 

The construction of two separated models instead of direct prediction of frustration index $f$ is conditioned due to several reasons. First, the Curie-Weiss and the N\'eel temperatures represent physically distinct magnetic observables and therefore have to be modeled independently rather than through direct prediction of the frustration index. The Curie-Weiss temperature reflects the overall balance of exchange interactions and its sensitivity to competing magnetic couplings and exchange topology. In contrast, the N\'eel temperature is primarily associated with stabilization of long-range magnetic order and depends more directly on dominant local exchange interactions and electronic configurations of magnetic ions. Consequently, independent prediction of $\Theta$ and $T_\mathrm{N}$ allows physically interpretable analysis of descriptor sensitivity and enables subsequent uncertainty-aware evaluation of the frustration index. 
In addition, the division of the ML workflows is caused by the fact that the experimental values of $\Theta$ and $T_\mathrm{N}$ originate from different sources in the literature and are available for different subsets of compounds.

Since NEMAD was constructed using large-language-model-assisted literature extraction, some degree of incompleteness, omission, or extraction error cannot be fully excluded. Therefore, additional data cleaning and consistency verification were performed prior to ML model development.The original NEMAD required extensive curation prior to ML. Material names were first standardized to recover chemically valid compositions. Since many entries contained additional comments describing synthesis conditions, particle size, isotope substitution, or sample morphology, these annotations were removed while preserving the chemical composition. The resulting formulas were subsequently parsed using the Pymatgen library~\cite{ong2013python}. Duplicate measurements were averaged, inconsistent records, and compositions with conflicting target values were additionally removed during the treatment. The final datasets contained $3942$ compounds for the prediction of $T_\mathrm{N}$ and $2378$ compounds for the prediction of $\Theta$. The distribution of the final sets for N\'eel and Curie-Weiss temperatures is shown in SFigure\,1a and 1b, respectively.

\subsection{Dataset subdivision}
\label{subdivition}
Since magnetic ordering temperatures arise from different microscopic exchange mechanisms, the dataset structure was further analyzed to determine whether chemically homogeneous subsets could improve predictive performance. The initial hypothesis considered two chemically relevant criteria: the presence of rare-earth (RE) elements and oxygen content. Rare-earth elements (Sc, Y, and the lanthanide series from La to Lu) introduce localized $4f$ magnetic moments and strong spin-orbit coupling effects~\cite{kanamori2006rare}, whereas oxygen-containing compounds typically exhibit superexchange-driven magnetic interactions~\cite{nqayi2025super,tatarchuk2024studying}. These fundamentally different magnetic mechanisms motivated subdivision of the datasets. To assess whether the resulting chemically defined subsets correspond to statistically distinct populations, non-parametric Kruskal-Wallis~\cite{kruskal1952use} and Mann-Whitney tests~\cite{mann1947test} were performed. The primary subdivision strategy therefore consisted of four chemically homogeneous classes: RE oxides, RE non-oxides, non-RE oxides, and non-RE non-oxides. The distribution of samples among these subclasses is shown in SFigure\,2a for the N\'eel temperature dataset and in SFigure\,2b for the Curie-Weiss temperature dataset. The resulting subset sizes were $849$, $1572$, $748$, and $785$ compounds for $T_\mathrm{N}$, and $567$, $844$, $439$, and $535$ structures for $\Theta$, respectively.

During subsequent model development, additional binary subdivision based on oxide/non-oxide and RE/non-RE chemistry were also investigated in order to evaluate whether simpler chemical partitioning provides a better compromise between chemical homogeneity and training-set size. For $T_\mathrm{N}$, the oxide and non-oxide subsets contain $1597$ and $2357$ compounds, respectively. For $\Theta$, the RE and non-RE subsets are of $1411$ and $974$ structures, respectively. Finally, each subset was processed independently using identical descriptor generation, train/test splitting, hyperparameter optimization, and evaluation protocols, in order to ensure that any differences in predictive performance originate solely from the way of subdivision.

\subsection{Descriptor generation}
\label{Subsec_descrip}
Several descriptor spaces of increasing complexity were initially evaluated to identify the most informative representation of the materials. These included (i) composition-based descriptors, (ii) composition supplemented with crystallographic symmetry information, and (iii) composition and symmetry supplemented by structural information. The predictive performance of these descriptor spaces was systematically compared during preliminary model development. Neither the inclusion of symmetry information nor additional structural features led to a measurable improvement in prediction accuracy. Therefore, all subsequent ML models were constructed using composition only descriptors. 

Composition-based descriptors were generated directly from chemical formulas using the Matminer library~\cite{ward2018matminer}. Three complementary descriptor families were employed. First, \textit{elemental-property statistics} were generated using the Magpie ElementProperty featurizer~\cite{ward2016general}. These descriptors characterize statistical distributions of elemental properties within each composition, including atomic number, atomic mass, electronegativity, covalent radius, Mendeleev number, melting temperature, valence-electron counts, ground-state magnetic moments, and numerous other elemental properties. Statistical measures including the mean, maximum, minimum, range, mode, and average deviation were calculated, resulting in $132$ descriptors.
Second, five \textit{stoichiometric descriptors} generated using the Stoichiometry featurizer were included to characterize compositional complexity and elemental fractions independently of elemental identity.
Finally, the ValenceOrbital featurizer was employed to calculate descriptors representing both \textit{the average numbers and fractional contributions of $s$-, $p$-, $d$-, and $f$-valence electrons} within each composition.

The resulting descriptor space consisted of $146$ numerical features for each compound and it was further used as the input for all ML models implemented in this work. In addition, relying solely on composition-based descriptors allows predictions for materials lacking experimentally resolved crystal structures, thereby considerably expanding the applicability of the proposed workflow for large-scale materials screening.

Although composition-only descriptors provided the best predictive performance among the tested descriptor sets, they do not explicitly describe exchange pathways, magnetic topology, orbital hybridization, or electronic structure. Consequently, some physically relevant information governing magnetic behavior is not directly captured by the model. The proposed workflow should therefore be regarded as a rapid screening approach rather than a substitute for detailed theoretical calculations or experimental investigations.

\subsection{Machine learning workflow}
\label{model}
Prediction of the Curie-Weiss and N\'eel temperatures was performed using the CatBoost gradient boosting regression algorithm~\cite{prokhorenkova2018catboost, veronika2018catboost}. CatBoost was selected because of its strong performance on heterogeneous tabular datasets, robustness against multicollinearity and noisy descriptors, and excellent predictive capability for descriptor-based materials-property prediction problems. Compared with deep-learning approaches, gradient boosting generally requires substantially fewer training samples and lower computational cost while often achieving comparable predictive accuracy for composition-based materials datasets~\cite{lazar2026beyond}. 

For each regression task, the dataset was randomly divided into training and test subsets using an $80/20$ split with a fixed random seed of $42$. Hyperparameter optimization was performed exclusively on the training subset using randomized search combined with ten-fold cross-validation. The optimization included tree depth, learning rate, number of boosting iterations, $L_2$ leaf regularization, subsampling ratio, random-strength regularization, and bagging-temperature parameters. The same optimization and training procedure was applied to both the complete datasets and all chemically defined subsets, ensuring direct comparability of the resulting models.

To improve prediction robustness and estimate predictive uncertainty, an ensemble of ten CatBoost models was constructed using ten-fold cross-validation on the training subset. Each model was trained on nine folds and used to predict the independent holdout test subset. The final prediction for each compound was obtained by averaging the predictions of all ensemble members, whereas the standard deviation among individual model predictions served as an estimate of predictive uncertainty.

To assess the consistency of the elemental composition between the training and independent holdout test subsets, the occurrence frequency of each element was calculated as the percentage of compounds containing that element in each subset. The resulting distributions for the $T_\mathrm{N}$ and $\Theta$ datasets are compared in SFigure\,3.

The standard deviation among ensemble predictions should be interpreted as a measure of model disagreement and therefore primarily reflects uncertainty arising from model variability and limitations of the available training data. It does not represent a fully calibrated probabilistic uncertainty and does not explicitly account for systematic experimental errors, descriptor incompleteness, or extrapolation beyond the chemical domain represented in the training dataset.

Further, the estimated uncertainty was subsequently incorporated into the screening procedure. Rather than ranking compounds solely according to the predicted frustration index, uncertainty estimates were used to identify high-confidence candidates and distinguish them from predictions associated with larger model uncertainty.

Model quality was evaluated using the mean absolute error (MAE), root mean square error (RMSE), and coefficient of determination ($R^2$). Three complementary evaluation protocols were employed. First, the average performance over the ten cross-validation folds was used to assess the stability of the training procedure. Second, out-of-fold (OOF) predictions were generated for the training subset, with each prediction obtained from a model that had not been trained on the corresponding sample. Finally, predictive performance was evaluated on an independent holdout test set excluded from hyperparameter optimization and model training. To further assess the dependence of predictive accuracy on chemical composition, errors on the independent test set were analyzed separately for four chemical classes defined by the presence or absence of rare-earth elements and oxygen, as described in Subsection \ref{subdivition}.

To interpret the learned relationships between descriptors and magnetic properties, SHapley Additive exPlanations (SHAP)~\cite{lundberg2017unified} were employed. Mean absolute SHAP values were used to rank descriptor importance, whereas SHAP summary plots were used to visualize both the magnitude and direction of descriptor contributions to the model predictions.

\subsection{Screening workflow}
\label{screening}
The trained ML models were subsequently employed for large-scale screening of candidate frustrated magnetic materials using the Materials Project database~\cite{horton2025accelerated,jain2013commentary}. 

To clarify the roles of the two databases in this workflow, NEMAD and the Materials Project serve complementary purposes. NEMAD primarily contains experimentally reported magnetic properties extracted from the literature and was therefore used as the source of experimental target values for model training. In contrast, the Materials Project provides computationally derived materials data obtained from first-principles calculations and a comprehensive infrastructure for materials screening and analysis. Thus, NEMAD defines the experimental training space, whereas the Materials Project provides the broader materials space in which the trained models are applied to identify previously unexplored candidates.

Because the developed workflow relies exclusively on composition-derived descriptors, all candidate compounds were operated using the same preprocessing and descriptor-generation procedure as employed for the training datasets, thereby ensuring complete consistency between model training and screening.

To avoid extrapolation beyond the chemical domain represented by the training data, the screening was restricted to material classes described in Subsection\,\ref{subdivition}. Duplicate entries were removed and only thermodynamically stable or metastable compounds satisfying $E_\mathrm{hull}<0.1$~eV/atom were retained. This criterion restricts the screening to compounds with realistic experimental synthesizability. In addition, materials with calculated band gaps larger than $5$\,eV were excluded to remove highly ionic wide-gap insulators that are unlikely to exhibit significant magnetic exchange interactions. More important, before screening analysis, all compositions present in the NEMAD database were removed from the Materials Project candidate pool. Consequently, all reported candidates represent previously unseen compositions that were not included in the experimental database used for model development. After all filtering steps, the screening dataset contained approximately $6.7\times10^4$ candidate compounds. 

Prediction uncertainty was estimated using the ensemble-learning strategy developed during the model construction. For each target property, predictions from all CatBoost ensemble members were averaged to obtain the final predicted value, whereas the standard deviation of the ensemble predictions was taken as the corresponding uncertainty measure:
\begin{equation}
T_\mathrm{N}
=
\left<T_\mathrm{N}^{(i)}\right>,
\qquad
\sigma_{T_\mathrm{N}}
=
\mathrm{std}
\left(
T_\mathrm{N}^{(i)}
\right),
\end{equation}

\begin{equation}
\Theta
=
\left<\Theta^{(i)}\right>,
\qquad
\sigma_{\Theta}
=
\mathrm{std}
\left(
\Theta^{(i)}
\right),
\end{equation}
where $i$ denotes individual ensemble members. The magnetic frustration index was subsequently calculated using Equation\,\ref{eq:f_index}.

Since both quantities are predicted independently, their uncertainties propagate to the calculated frustration index. Assuming independent prediction errors, the relative uncertainty was estimated using the standard propagation-of-errors formalism\,(SEquation\,1)~\cite{taylor1997introduction}.

To suppress false candidates originating from uncertain predictions, an uncertainty-aware conservative frustration index was additionally introduced~\cite{de2021conservatism}:
\begin{equation}
f_{\mathrm{cons}}
=
\frac
{|\Theta|-\sigma_{\Theta}}
{T_\mathrm{N}+\sigma_{T_\mathrm{N}}},
\label{eq:fcons}
\end{equation}
which suggests a conservative estimate of the $f$ by explicitly accounting for the prediction uncertainties of both temperatures. %Compared with 
The conservative frustration index is a heuristic uncertainty-penalized screening score rather than a statistically calibrated lower confidence bound.
Consequently, only compounds exhibiting both large predicted frustration and sufficiently confident temperature predictions retain high values of $f_{\mathrm{cons}}$. 

Finally, all screened compounds were ranked according to $f_{\mathrm{cons}}$, and materials with the highest conservative frustration indices were selected as the most promising candidates for further investigation.

\section{Results and discussion}
\label{results}
\subsection{Machine learning model performance and descriptor analysis}
CatBoost regression models were first implemented for the complete $T_\mathrm{N}$ and $\Theta$ datasets using the composition-based descriptor space described in Section\,\ref{Subsec_descrip}. Further, prediction performance was evaluated. The corresponding metrics obtained for the test set for complete set of materials are presented in Table\,\ref{tab:subdivision_summary}, whereas full set of metrics, including CV and OOF are presented in STable\,1 of the Supplementary Materials, confirming model stability.
\begin{table}[ht]
\centering
\caption{Comparison of the global set model performance with the best-performing subdivision strategies on the independent test set.}
\label{tab:subdivision_summary}
\small
\begin{tabular}{llcccc}
\hline
Target & Metric & Full dataset & Binary subdivision & Four-class subdivision \\
\hline

\multirow{3}{*}{$T_\mathrm{N}$}
& MAE  & 43.00 & 54.27 (Oxides) & 38.51 (RE oxide) \\
& RMSE & 88.07 & 95.05 (Oxides) & 67.71 (RE oxide) \\
& $R^2$ & 0.73 & 0.76 (Oxides) & 0.83 (RE oxide) \\

\hline

\multirow{3}{*}{$\Theta$}
& MAE   & 63.09 & 34.71 (RE) & 27.55 (RE nonoxide) \\
& RMSE  &127.85 & 83.98 (RE) & 58.51 (RE nonoxide) \\
& $R^2$ &0.63   & 0.72 (RE)  & 0.59 (RE nonoxide) \\

\hline
\end{tabular}
\end{table}

For the complete dataset, prediction of the N\'eel temperature achieved a test coefficient of determination of $R^2=0.73$, together with $R^2_\mathrm{CV}=0.67$ and $R^2_\mathrm{OOF}=0.67$. The corresponding mean absolute error on the independent test set was $43.0$\,K. Prediction of the Curie-Weiss temperature reached a test coefficient of determination of $R^2=0.63$, with $R^2_\mathrm{CV}=0.60$ and $R^2_\mathrm{OOF}=0.60$. These results indicate that a substantial fraction of magnetic interaction information is already present in composition elemental statistics and electronic descriptors.

Overall, the obtained prediction accuracy is moderate, however, it is expected for the experimental data. Both $\Theta$ and $T_\mathrm{N}$ are governed by a complex interplay of exchange interactions, magnetic anisotropy, electronic structure, and long-range magnetic correlations. In addition, experimental values collected from the literature are affected by variations in sample quality, synthesis conditions, defects, non-stoichiometry, and measurement protocols. Furthermore, the descriptor space does not explicitly include quantities such as exchange constants, or electronic density of states, which also contribute to the observed magnetic behavior. Similar moderate prediction accuracies of $R^2 =0.66$ have recently been reported for CatBoost prediction of Curie temperatures in chemically diverse materials datasets derived from literature, using elemental descriptors~\cite{orang2026curie}. However, the authors showed the possibility to improve $R^2$ up to $0.87$ using manually curated database without restriction of the features space.  
Additionally, representative parity plots for the final models are shown in SFigure\,4, revealing that in both cases, most part of compounds are distributed close to the diagonal line, with expected deviations.

\begin{figure}[ht!]
\centering
\begin{subfigure}{0.49\linewidth}
  \centering
  \includegraphics[width=\linewidth]{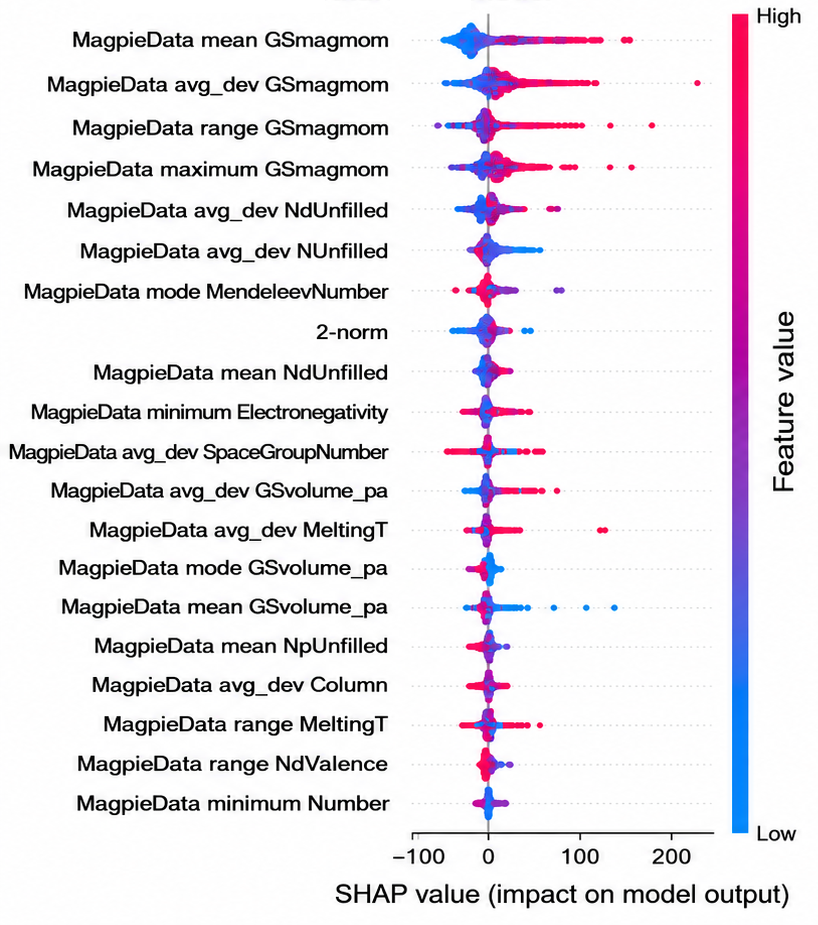}
  \caption{}
  \label{fig:importance_SHAP-a}
\end{subfigure}
\hfill
\begin{subfigure}{0.49\linewidth}
  \centering
  \includegraphics[width=\linewidth]{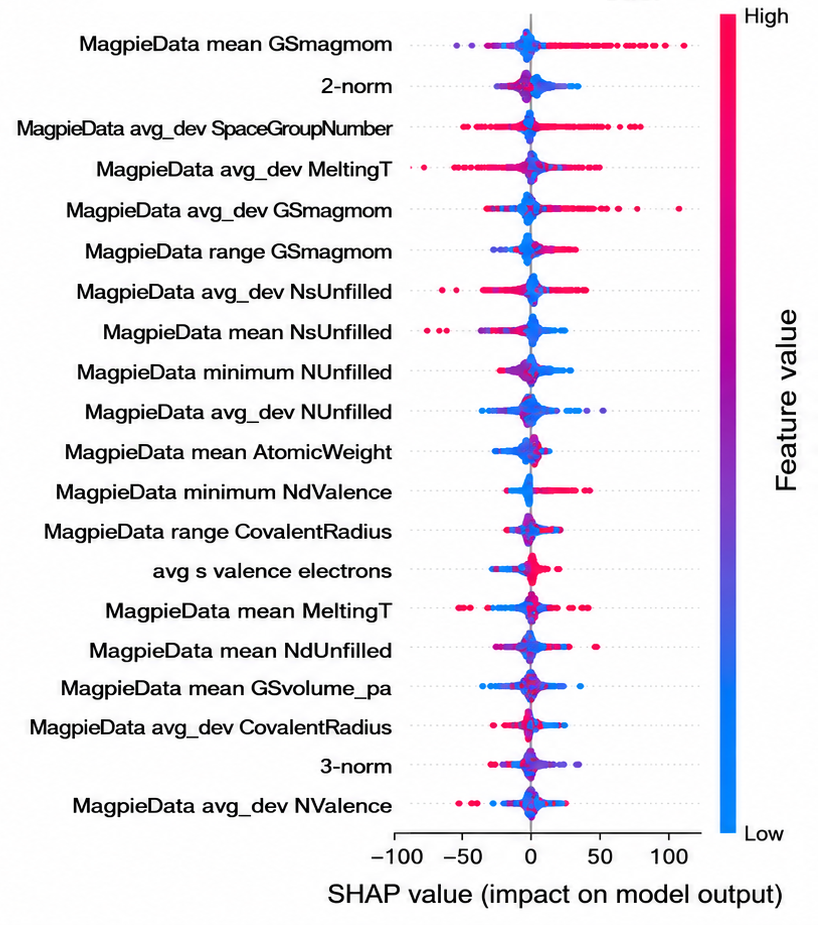}
  \caption{} 
  \label{fig:importance_SHAP-b}
\end{subfigure}
\caption{SHAP beeswarm plots showing the contribution of the twenty most important descriptors to the prediction of (a) N\'eel temperature and (b) Curie–Weiss temperature.}
\label{fig:importance_SHAP}
\end{figure}

To obtain physical interpretation of the learned models, SHAP analysis was performed for both target properties separately. Prediction of the N\'eel temperature is governed primarily by descriptors associated with magnetic properties of constituent elements (Figure\,\ref{fig:importance_SHAP-a}). 
It was found that the most influential feature is the mean ground-state magnetic moment (GSmagmom), followed by its average deviation, range, and maximum value. This result is physically expected because AFM ordering temperatures are determined by the strength of magnetic exchange interactions, which depend strongly on the magnitude and distribution of local magnetic moments. The beeswarm plot further demonstrates that larger values of these magnetic descriptors (red points) predominantly produce positive SHAP values, indicating a tendency toward higher predicted $T_\mathrm{N}$. Conversely, compounds with small average magnetic moments generally contribute negatively to the prediction.

Besides magnetic descriptors, the model also assigns considerable importance to descriptors related to the electronic structure, including the variation of the numbers of unfilled  $d$- and  $f$-shell electrons (NdUnfilled and NUnfilled). These descriptors reflect differences in partially filled electronic shells, which strongly influence exchange interactions in transition-metal and rare-earth compounds. 

Additional contributions arise from crystal-chemical descriptors, such as the Mendeleev number, electronegativity, atomic volume, melting temperature, and space-group statistics. Their lower but non-negligible importance suggests that structural and chemical environments modulate the magnetic interactions rather than determining them directly.

Overall, the SHAP analysis indicates that the model captures physically meaningful relationships, where magnetic descriptors dominate the prediction while electronic and structural descriptors provide secondary corrections.

The SHAP analysis for the Curie–Weiss temperature (Figure\,\ref{fig:importance_SHAP-b}) reveals a similar but more diverse descriptor hierarchy. Again, the mean ground-state magnetic moment is identified as the most influential feature, confirming that the strength of local magnetic moments is the primary factor controlling magnetic interaction scales. Unlike the N\'eel-temperature model, however, the Curie–Weiss prediction depends more strongly on descriptors describing the crystal structure and chemical diversity. Among the most important variables are the 2-norm stoichiometric descriptor, the average deviation of the space-group number, melting temperature, and several descriptors related to unfilled valence orbitals and atomic radii.

Additionally, the beeswarm distribution shows that many descriptors exhibit both positive and negative SHAP contributions over different value ranges, indicating pronounced nonlinear behavior. Such behavior is expected because the Curie–Weiss temperature reflects the net exchange interaction, which depends on a delicate balance between competing FM and AFM exchange pathways rather than solely on magnetic moment magnitude.
Finally, compared with the $T_\mathrm{N}$ model, the importance of descriptors for the Curie–Weiss temperature prediction is distributed more evenly across multiple descriptors, suggesting that predicting $\Theta$ requires simultaneous consideration of magnetic, structural, and chemical characteristics.

\subsection{Statistical analysis of chemically defined subsets}
\begin{figure}[htbp]
\centering
\begin{subfigure}{0.7\linewidth}
  \centering
  \includegraphics[width=\linewidth]{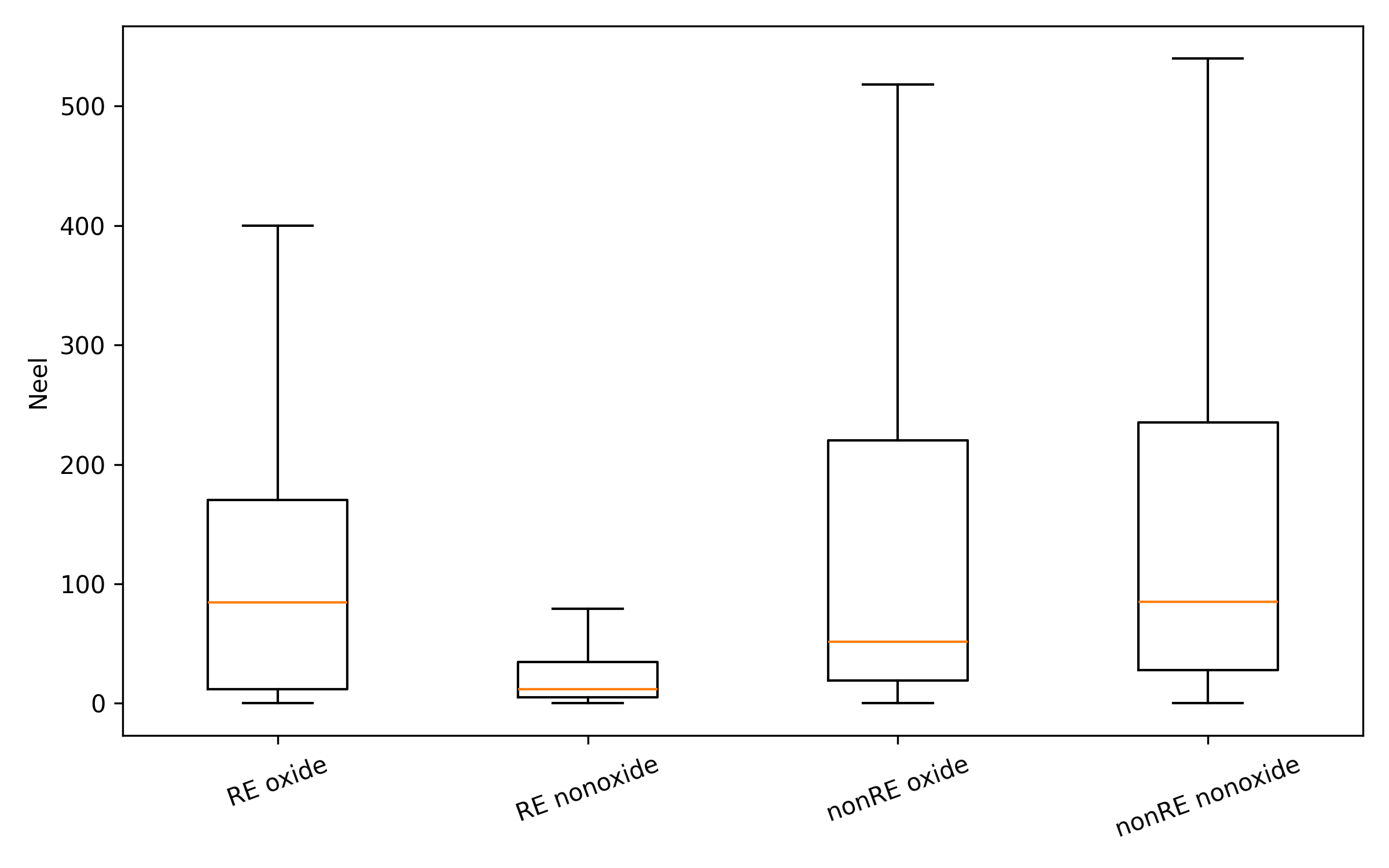}
  \caption{}
  \label{fig:groups-a}
\end{subfigure}
\hfill
\begin{subfigure}{0.7\linewidth}
  \centering
  \includegraphics[width=\linewidth]{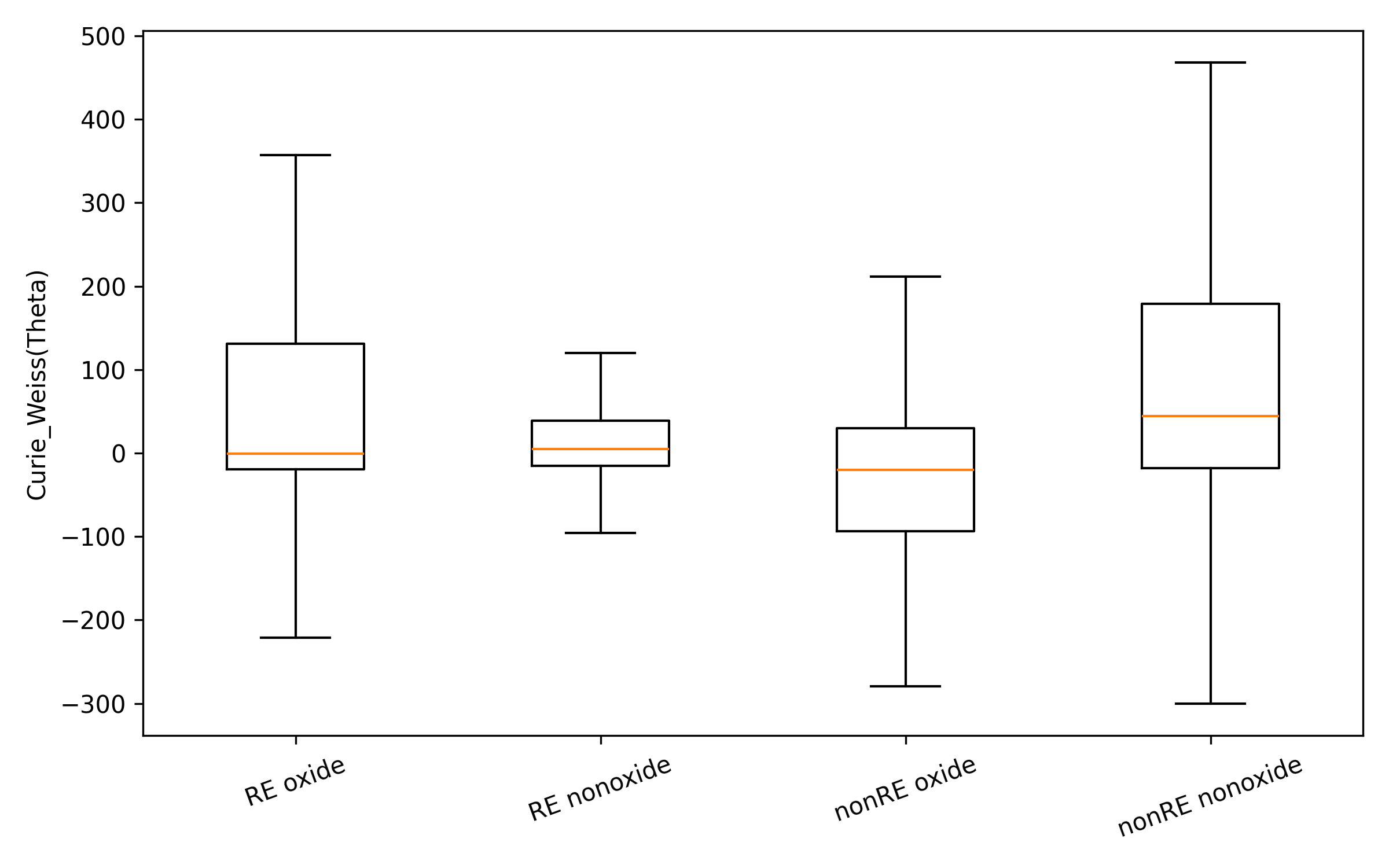}
  \caption{} 
  \label{fig:groups-b}
\end{subfigure}
\caption{Distribution of (a) N\'eel temperature and (b) Curie-Weiss temperature across chemically defined subclasses.}
\label{fig:groups}
\end{figure}

Before comparing different dataset subdivision strategies, the statistical distribution of the sets was analyzed to determine whether the proposed chemical subclasses correspond to distinct populations. Figure\,\ref{fig:groups-a} and Figure\,\ref{fig:groups-b} present the distributions of the N\'eel and Curie-Weiss temperatures for the four chemically defined material classes. 
The distributions of both magnetic properties differ significantly across the four chemical subclasses. For $T_\mathrm{N}$, the strongest statistical separation is associated with oxygen content, whereas for $\Theta$ the dominant differences arise from the presence of rare-earth elements. Non-parametric Kruskal–Wallis tests rejected the hypothesis that all subclasses originate from the same population ($p \ll 0.001$), and pairwise Mann–Whitney tests confirmed these trends (STables\,3-4). These observations demonstrate that the proposed subclasses represent statistically distinct populations. Therefore, the development of chemically divided ML models provides a physical justification.

\subsection{Effect of dataset subdivision on model performance}
A statistical analysis (STable\,2) showed that the four-class subdivision is physically meaningful but not always optimal for ML. Therefore, binary partitions were additionally explored: oxide/non-oxide for $T_\mathrm{N}$ and RE/non-RE for $\Theta$. For $T_\mathrm{N}$, the oxide model ($R^2=0.76$) slightly outperformed the global model ($R^2=0.73$), although the best four-class model remained superior. In contrast, for $\Theta$, the RE/non-RE subdivision proved most effective, with both models ($R^2=0.72$ and $R^2=0.68$) outperforming the global and all four-class models. These results indicate that the optimal subdivision strategy is target dependent: prediction of the N\'eel temperature benefits from maximizing chemical homogeneity, whereas prediction of the Curie-Weiss temperature is optimized by the simpler RE/non-RE partitioning. Accordingly, the best-performing models were selected for large-scale screening of candidate frustrated magnetic materials.
The corresponding predictive performance is summarized in STable\,5, along with detailed CV and OOF metrics.

\begin{figure}[htbp]
\centering
\begin{subfigure}{0.49\linewidth}
\centering
\includegraphics[width=\linewidth]{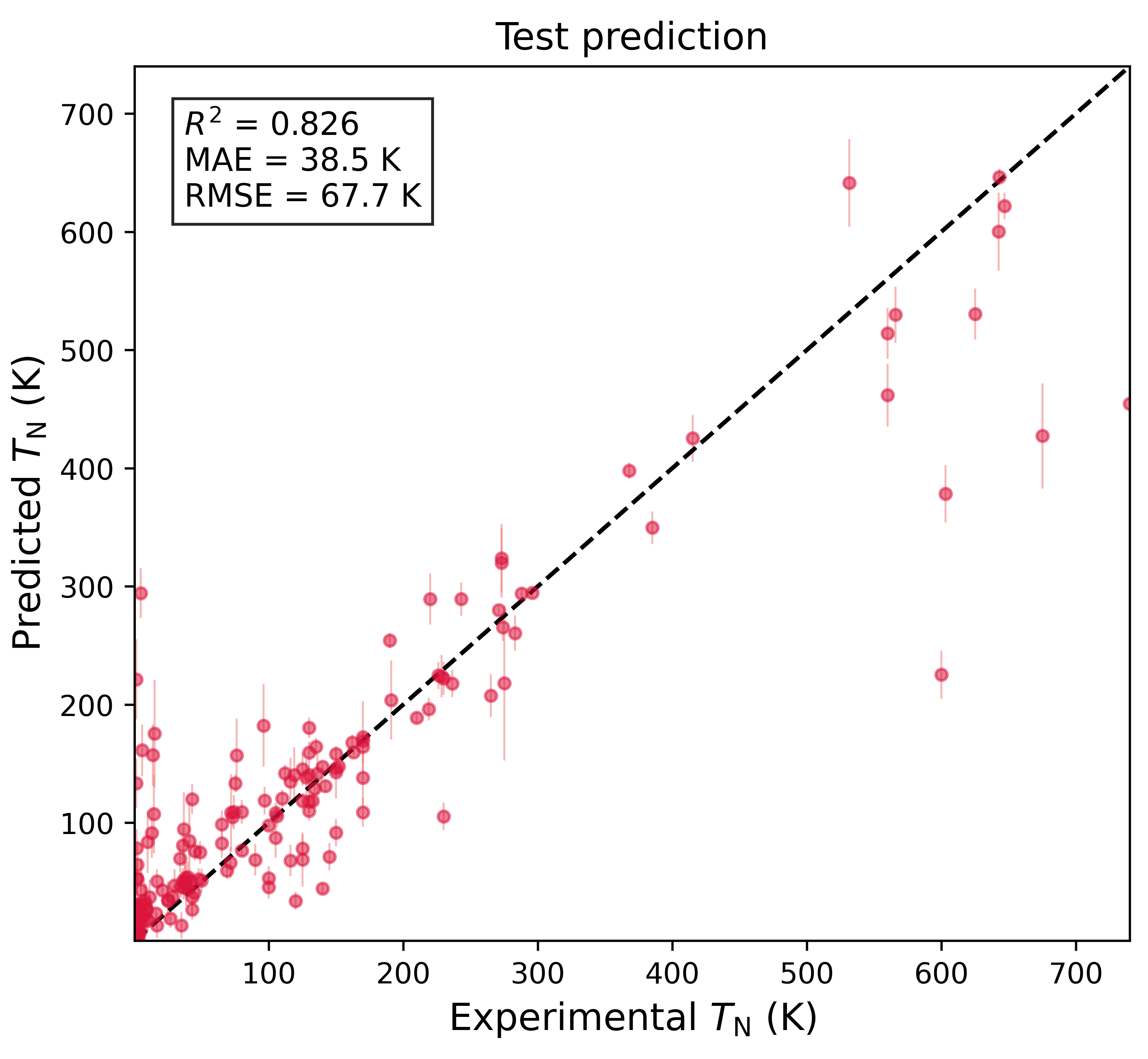}
\caption{}
\label{fig}
\end{subfigure}
\hfill
\begin{subfigure}{0.49\linewidth}
\centering
\includegraphics[width=\linewidth]{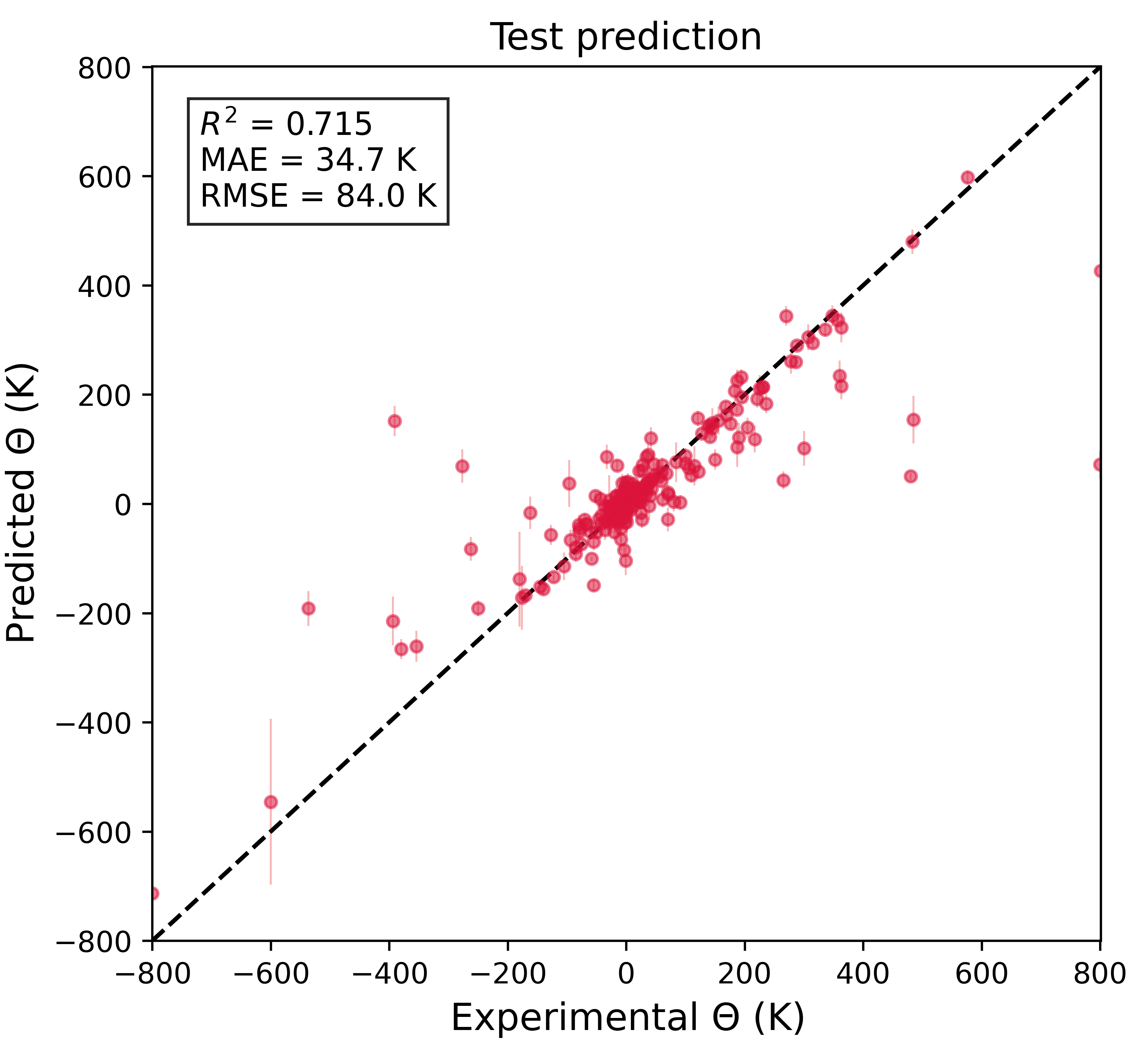}
\caption{}
\label{fig}
\end{subfigure}
\caption{Parity plots of (a) N\'eel temperature and (b) Curie-Weiss temperature for the best chemical subdivision strategies.}
\label{figparity_best}
\end{figure}

Parity plots for the best subdivision strategy for the N\'eel and Curie-Weiss temperatures are presented in Figure\,\ref{figparity_best}a and \ref{figparity_best}b, correspondingly.

The different training subsets used for the $T_\mathrm{N}$ and $\Theta$ models were further examined by comparing their element-level contribution profiles (Figure\,\ref{fig}). The profiles are broadly similar for the most prominent elements, particularly O and Mn, while several elements show target-dependent differences. For example, Fe has a larger contribution in the $T_\mathrm{N}$ model, whereas Co, Gd, and La contribute more strongly to the $\Theta$ model. These differences may reflect both the distinct chemical compositions of the training subsets and the different physical quantities represented by the two targets, rather than indicating an intrinsic elemental bias of the models.

\begin{figure}[htbp]
\centering
\includegraphics[width=0.85\linewidth]{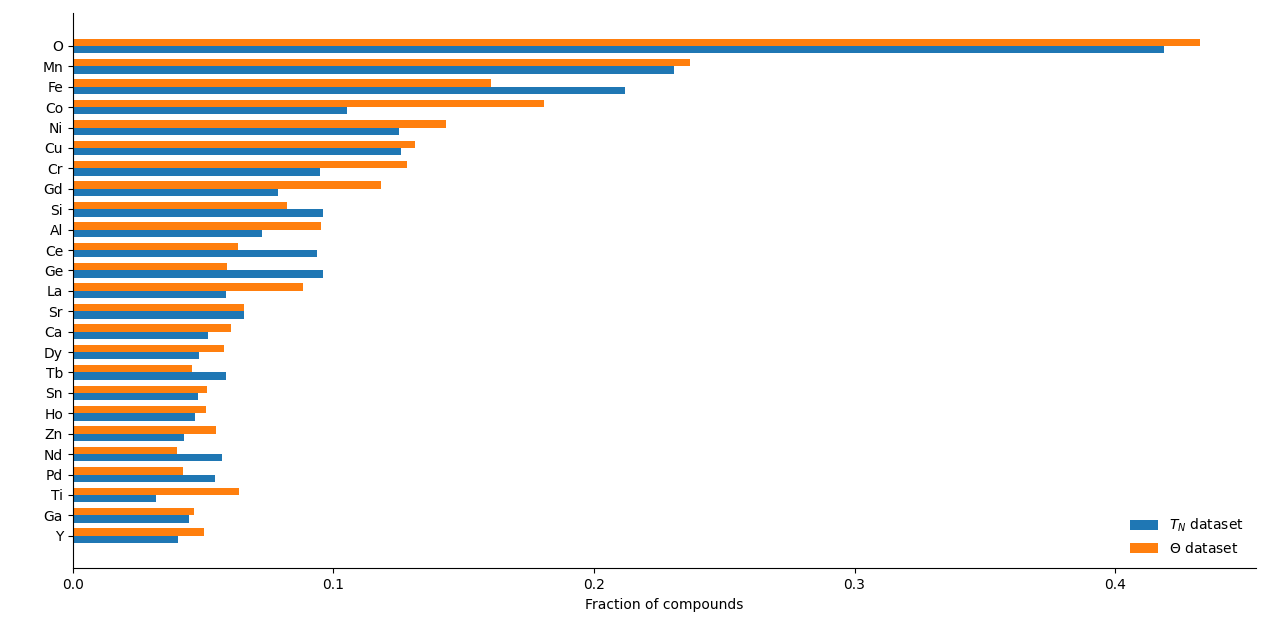}
\caption{Element-level contribution profiles for the $T_\mathrm{N}$ and $\Theta$ models.}
\label{fig}
\end{figure}

\begin{figure}[htbp]
\centering

\begin{minipage}[t]{0.48\linewidth}
    \centering
    \includegraphics[width=\linewidth]{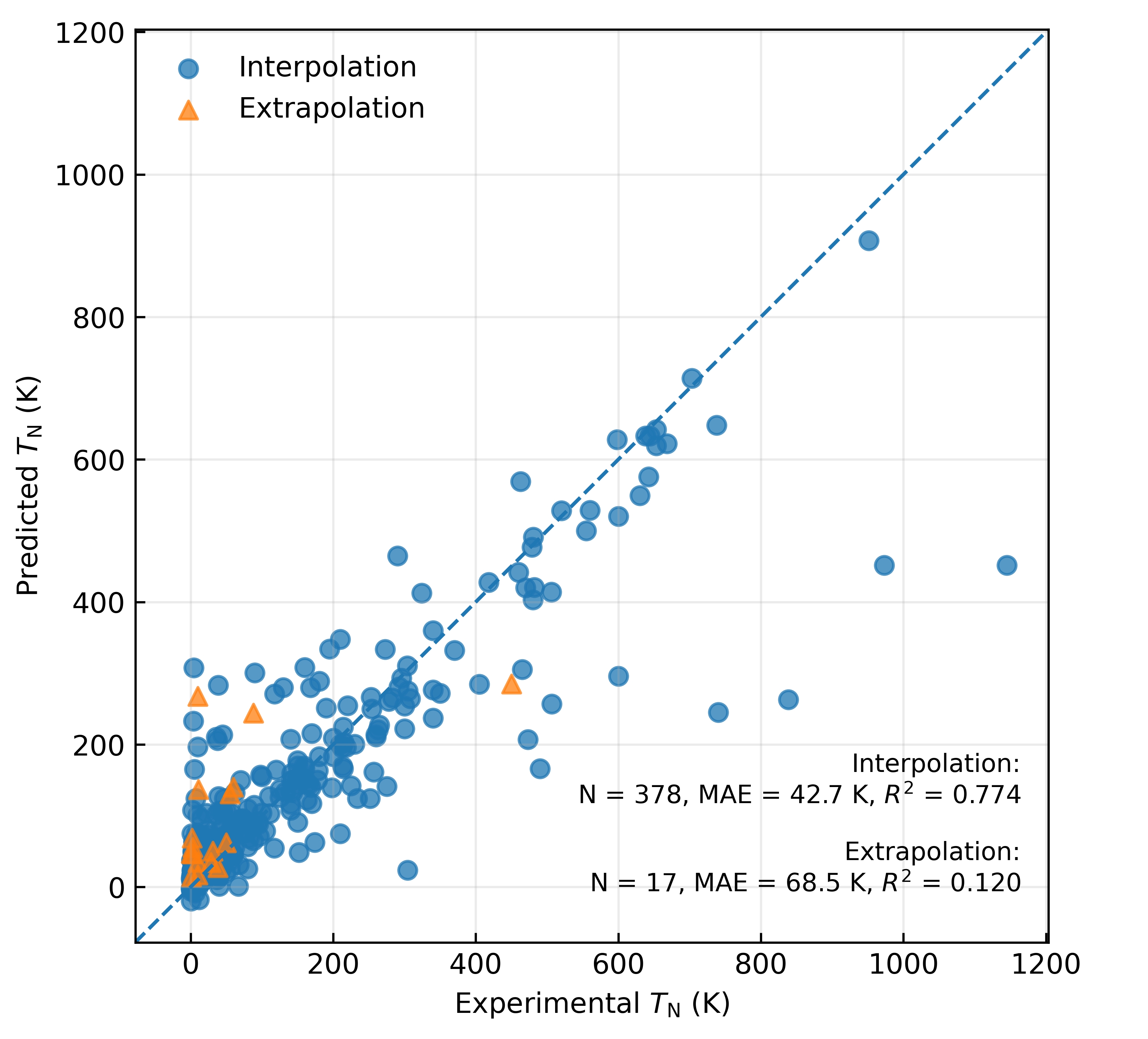}
    \centerline{(a)}
\end{minipage}
\hfill
\begin{minipage}[t]{0.48\linewidth}
    \centering
    \includegraphics[width=\linewidth]{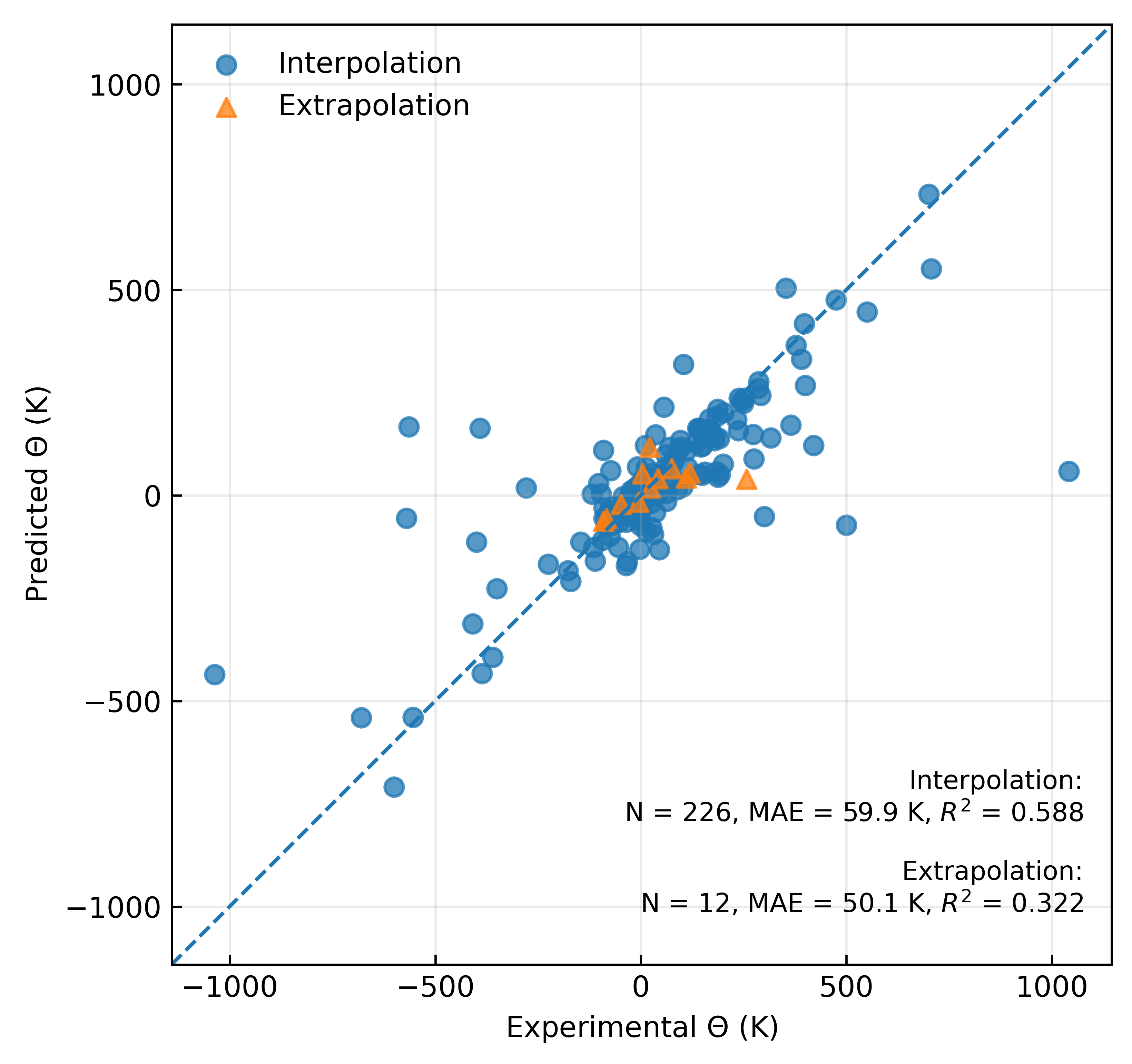}
    \centerline{(b)}
\end{minipage}

\caption{Parity plots showing model performance for interpolation and chemical extrapolation cases for (a) $T_\mathrm{N}$ and (b) $\Theta$. The interpolation/extrapolation boundary was defined using the 95th percentile of the training-set nearest-neighbor chemical distance in the standardized descriptor space.}
\label{fig:interpolation_extrapolation}
\end{figure}

The sensitivity of the models to chemical extrapolation was additionally evaluated (Figure\,\ref{fig:interpolation_extrapolation}). Using the $95$th percentile of the training-set nearest-neighbor distance in the standardized descriptor space as the extrapolation threshold, 4.3\% of the $T_\mathrm{N}$ test samples and 5.0\% of the $\Theta$ test samples in the first cross-validation fold were classified as extrapolation cases. For $T_\mathrm{N}$, the interpolation subset yielded MAE = 42.7,K and $R^2 = 0.774$, compared with MAE = 68.5\,K and $R^2 = 0.120$ for extrapolation. For $\Theta$, the corresponding values were MAE = 59.9\,K and $R^2 = 0.588$ for interpolation and MAE = 50.1\,K and $R^2 = 0.322$ for extrapolation. Thus, both models show reduced explained variance for chemically extrapolated samples, while the effect on absolute error is less systematic for $\Theta$.

\subsection{Screening of frustrated magnetic candidates}
\label{sub:screen}
As described in the previous section, the final screening models were selected based on the optimal balance between predictive accuracy and chemical transferability. These models were subsequently applied for large-scale identification of frustrated magnetic candidates among the selected materials space. It is worth noting that the selected models retain predictive capability beyond their original training subsets. The cross-class evaluation demonstrated satisfactory transferability, particularly for oxide systems, indicating that the screening results are not restricted to the specific chemical classes used during model optimization. 

\begin{figure}[h!]
    \centering
    \includegraphics[width=0.8\linewidth]{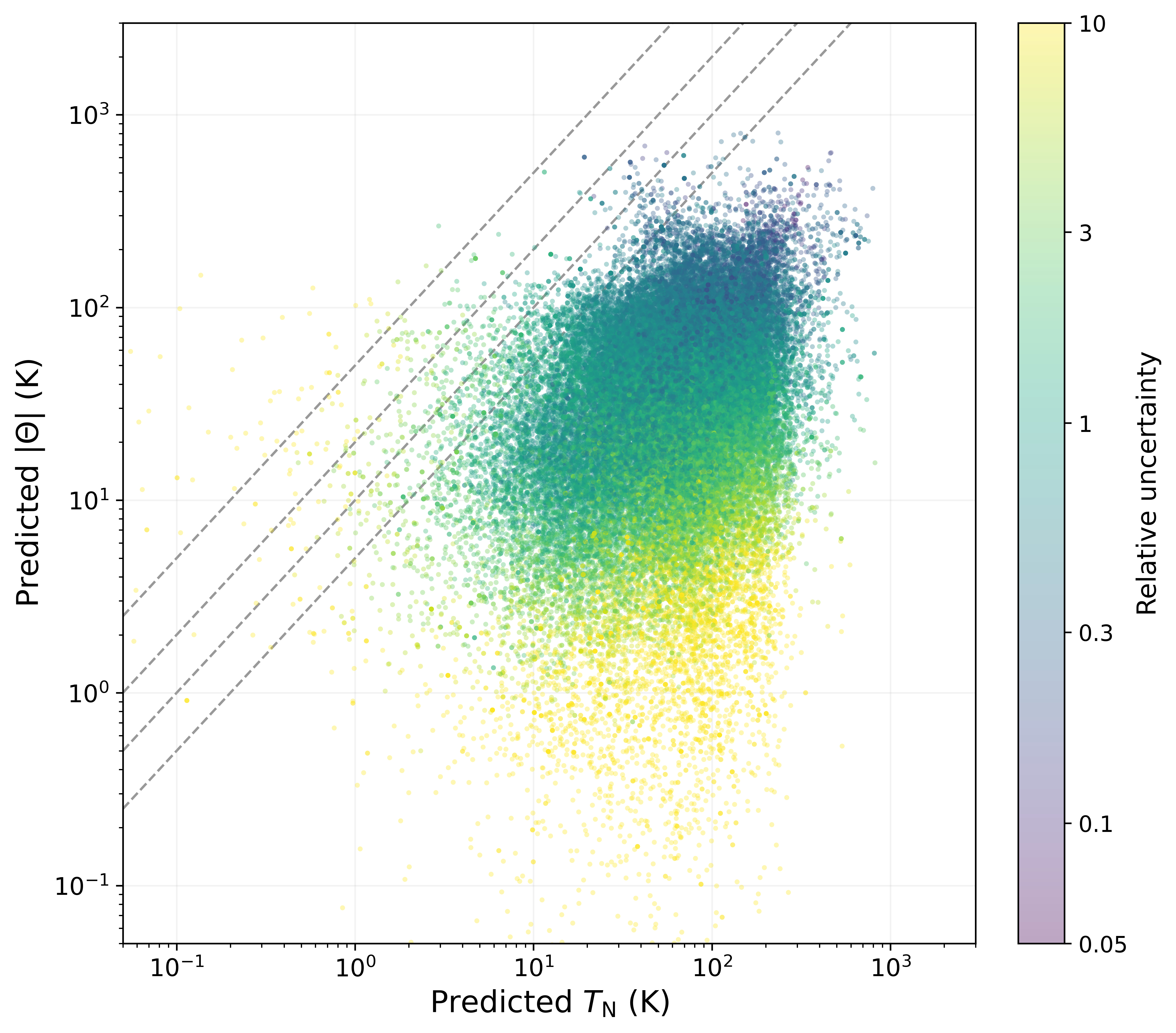}
    \caption{Predicted frustration map for screened candidate materials. Dashed lines correspond to constant frustration levels of $f=5$, $f=10$, $f=20$, and $f=50$. The color scale represents relative prediction uncertainty estimated from ensemble variance.}
    \label{fig:frustration_map}
\end{figure}

The resulting predicted frustration map is shown in Figure\,\ref{fig:frustration_map}. 
As seen, the majority of screened compounds are located in the weakly frustrated or non-frustrated region, whereas only a relatively small fraction exhibits frustration indices exceeding the commonly adopted threshold of $f=5$. Additionally, the color scale in Figure\,\ref{fig:frustration_map} represents relative prediction uncertainty estimated from ensemble model variance. Although several compounds exhibit very large predicted frustration indices, some of these predictions are accompanied by substantial uncertainty, indicating that a high predicted frustration index alone is insufficient for reliable candidate identification.

To improve the robustness of the screening procedure, the final ranking was based on the conservative frustration index $f_{\mathrm{cons}}$, defined in Equation,\ref{eq:fcons}, rather than on the conventional frustration index alone. The distribution of the relative prediction uncertainty for all screened compounds is presented in SFigure\,5. To assess the sensitivity of the screening results to the relative-uncertainty threshold, we considered three threshold values ($0.2$, $0.3$, and $0.5$), while keeping the criterion $f_{\mathrm{cons}} \geq 5$ fixed (SFigure\,6). These thresholds yielded $22$, $57$, and $125$ candidate entries before deduplication, respectively. After chemical-formula normalization and removal of duplicate formulas, the corresponding numbers of unique candidates were $17$, $36$, and $70$, respectively. The threshold of $0.3$ was retained as a practical compromise between limiting prediction uncertainty and maintaining a sufficient number of screening candidates, yielding $36$ unique candidates for further consideration.

\begin{figure}[h!]
    \centering
    \includegraphics[width=0.8\linewidth]{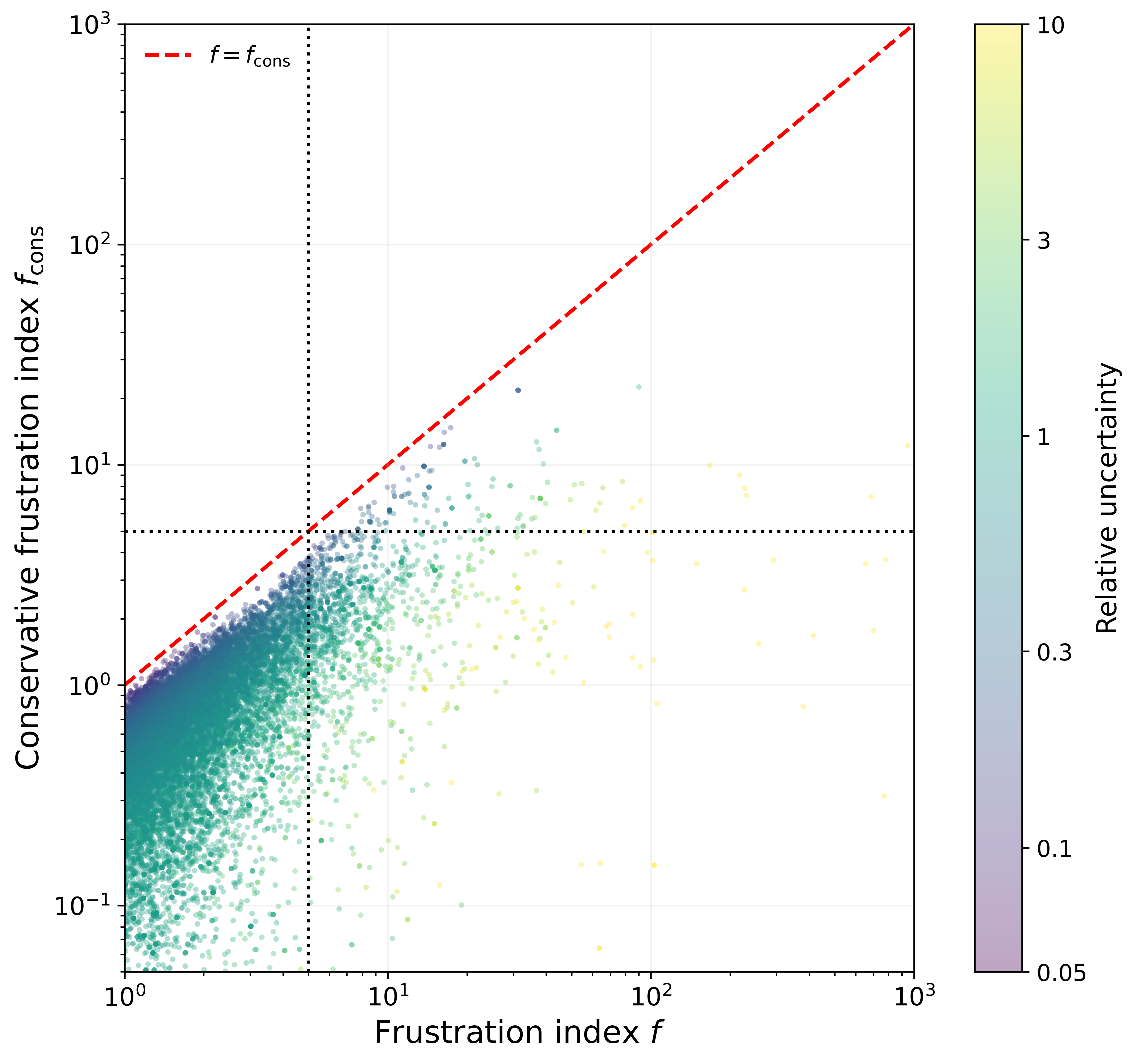}
    \caption{Relationship between the conventional frustration index ($f$) and the uncertainty-aware conservative frustration index ($f_{\mathrm{cons}}$).}
    \label{fig:uncertainty_effect}
\end{figure}

Figure\,\ref{fig:uncertainty_effect} illustrates the influence of uncertainty correction on the predicted frustration indices.
The red dashed line corresponds to the ideal case $f_{\mathrm{cons}} = f$, for which uncertainty has no effect on the predicted frustration. The black dotted lines indicate the frustration threshold of $f=5$. The color scale represents the relative prediction uncertainty estimated from the ensemble variance. As expected, all compounds are located below red line because the conservative formulation simultaneously decreases the predicted Curie-Weiss temperature and increases the predicted magnetic ordering temperature within their respective uncertainty intervals. The deviation from the diagonal increases with the relative prediction uncertainty, demonstrating that compounds associated with less reliable predictions undergo a stronger reduction of the frustration index. Consequently, a considerable fraction of materials exhibiting $f > 5$ no longer satisfies the conservative screening criterion after uncertainty correction. In contrast, compounds remaining above the horizontal threshold $f_{\mathrm{cons}} = 5$ combine both a large predicted frustration index and sufficiently reliable ML predictions, making them the most promising candidates for subsequent experimental validation.

In addition, SFigure\,7 compares the fraction of screened compounds exceeding different frustration thresholds when using the conventional frustration index and the conservative frustration index. The conservative formulation systematically reduces the number of retained candidates, with the difference becoming increasingly pronounced at larger frustration thresholds. This behavior demonstrates that uncertainty-aware correction effectively suppresses highly uncertain extreme predictions while preserving the general screening trends.

\subsection{Chemical trends in screened candidates}
\label{chemistry}
\begin{figure}[htbp]
\centering
\begin{subfigure}{0.6\linewidth}
  \centering
  \includegraphics[width=\linewidth]{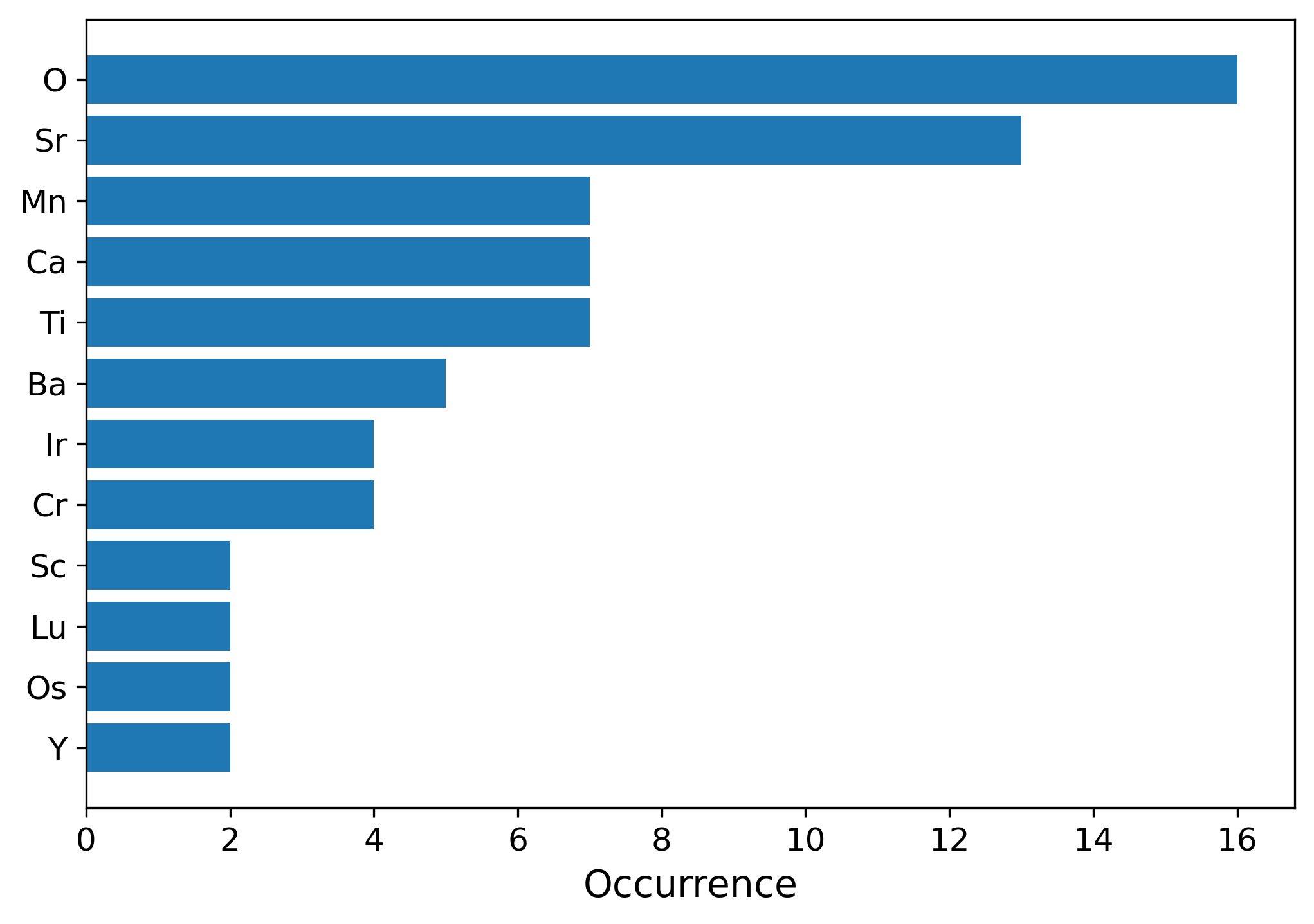}
  \caption{}
  \label{fig:chemistry-a}
\end{subfigure}
\hfill
\begin{subfigure}{0.65\linewidth}
  \centering
  \includegraphics[width=\linewidth]{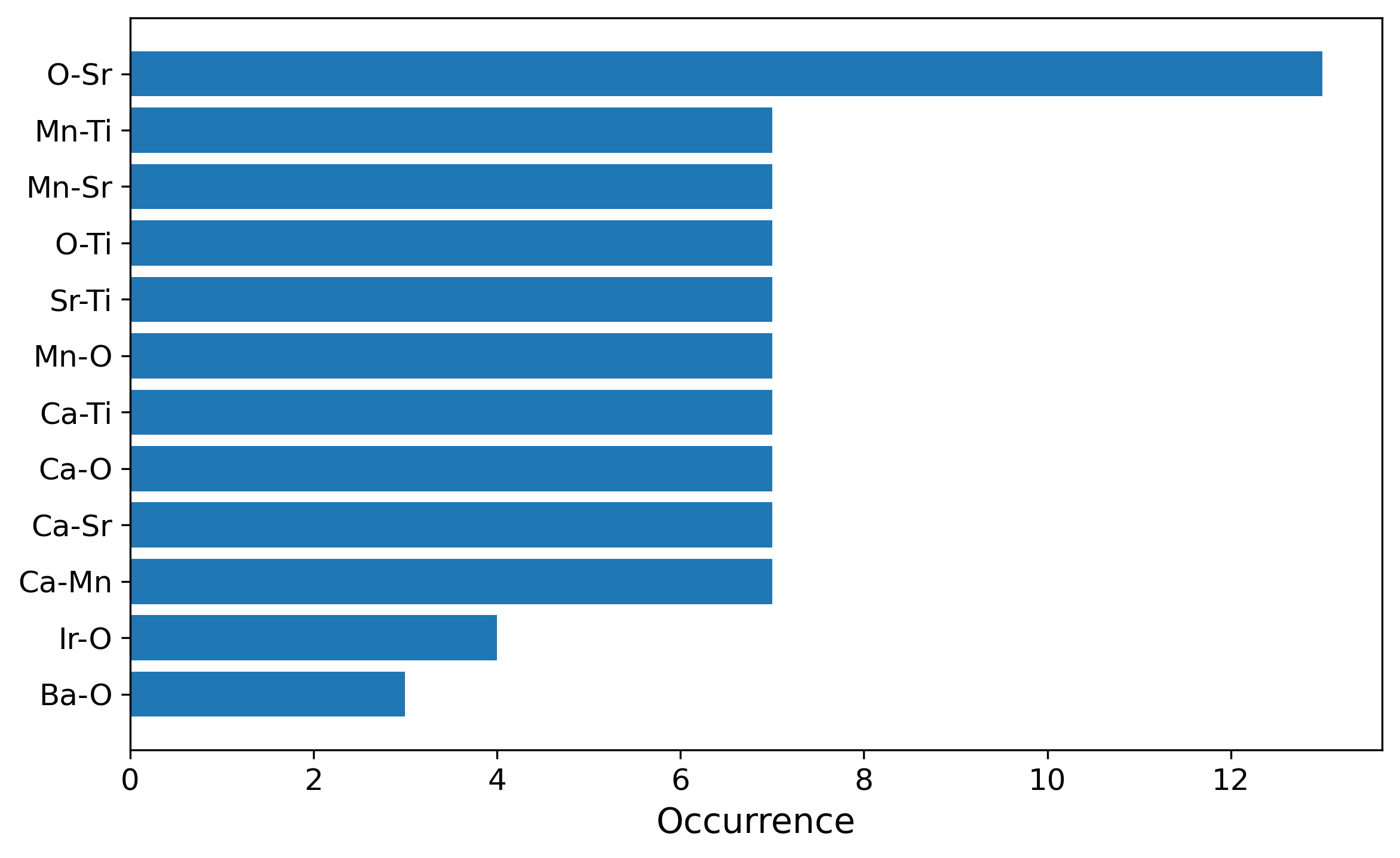}
  \caption{} 
  \label{fig:chemistry-b}
\end{subfigure}
\caption{Chemical statistics of the final screened candidate set after uncertainty-aware conservative filtering. (a) Frequency of individual chemical elements. (b) Most frequently occurring elemental pairs.}
\label{fig:chemistry}        
\end{figure}

To investigate the chemical and structural characteristics of the screened materials, the elemental composition, repeating elemental combinations, and crystallographic symmetry were analyzed. Structural information was obtained from the Materials Project database. The chemical statistics of the final candidate set are summarized in Figure\,\ref{fig:chemistry}. 
The elemental-frequency analysis shown in Figure\,\ref{fig:chemistry-a} reveals a clear predominance of oxygen, followed by Sr, Mn, Ca, and Ti. These five elements constitute the chemical structure of the final screened materials and indicate that the highest-confidence candidates are primarily oxide compounds based on perovskite and related crystal chemistries. In contrast, only a few representatives containing heavier transition metals, such as Ir, Os, and Cr, remain after conservative filtering.

The predominance of oxide systems is consistent with the fact that transition-metal oxides are experimentally well known frustrated magnetic materials, where competing exchange interactions originate from interconnected magnetic sublattices, superexchange pathways, and structural distortions~\cite{ramirez1994strongly,gardner2010magnetic}.

A similar trend is observed in the elemental-pair analysis (Figure\,\ref{fig:chemistry-b}). The most frequent atomic pair is O-Sr, followed by Mn-Ti, Mn-Sr, O-Ti, Sr-Ti, Mn-O, Ca-Ti, Ca-O, Ca-Sr, and Ca-Mn. Nearly all dominant combinations belong to the same Sr-Ca-Ti-Mn oxide chemistry, demonstrating that the screening procedure repeatedly converges toward a limited number of chemically meaningful structural motifs instead of selecting chemically random compositions.

\begin{figure}[htbp]
\centering
\includegraphics[width=0.65\linewidth]{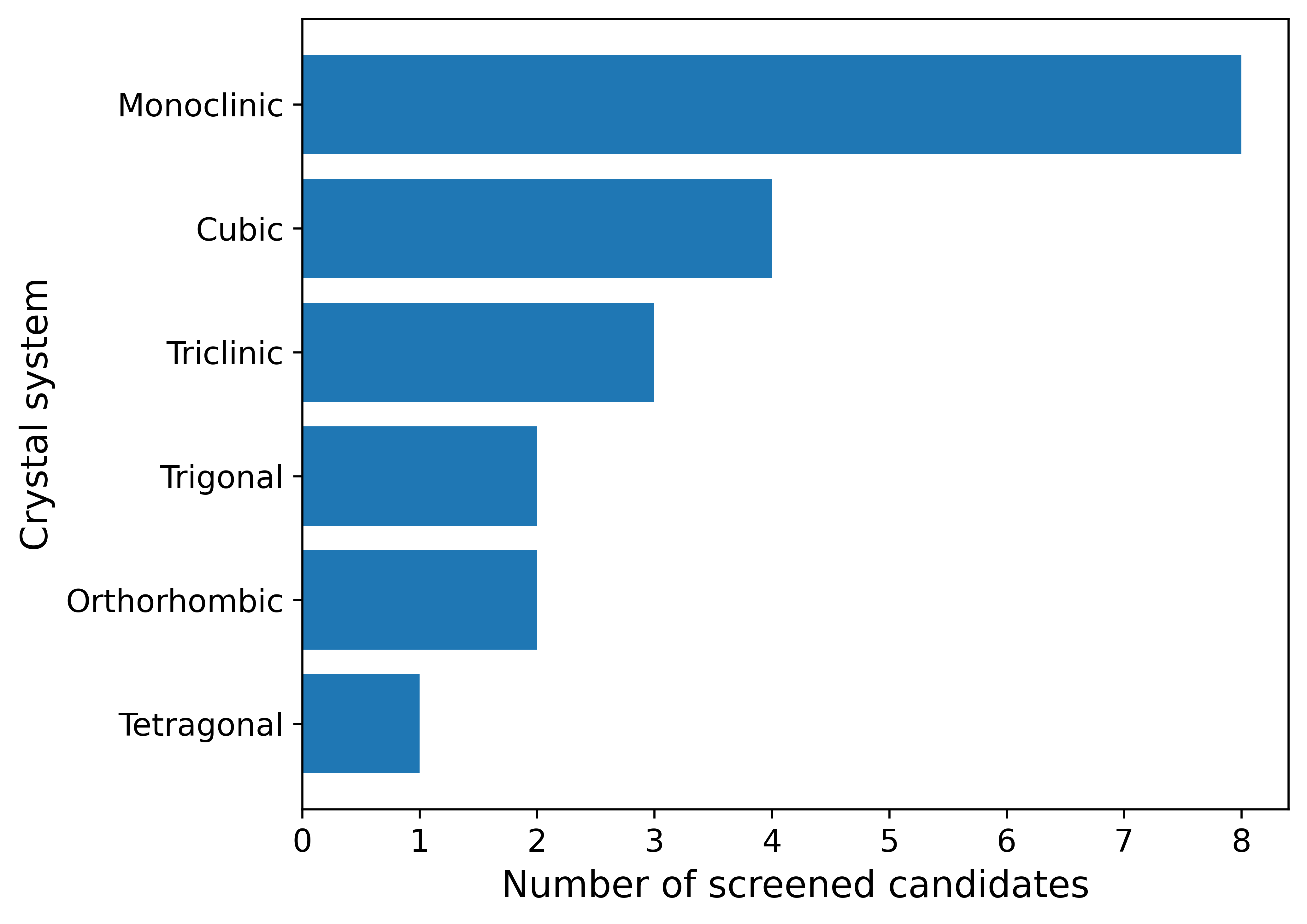}
\caption{Distribution of crystal systems among the final screened candidate materials after uncertainty-aware conservative filtering.}
\label{fig:space_group_classes}
\end{figure}

The crystallographic statistics provide an additional structural perspective on the screened candidates (Figure\,\ref{fig:space_group_classes}).
The final candidates is dominated by monoclinic structures (8 compounds), followed by cubic (4 compounds), triclinic (3 compounds), trigonal (2 compounds), orthorhombic (2 compounds), and tetragonal (1 compound). The large fraction of monoclinic compounds mainly reflects distorted oxide structures, where symmetry lowering produces multiple inequivalent magnetic exchange pathways that may enhance magnetic frustration~\cite{goto2021geometrical,duffort2013lifting}. Cubic structures primarily correspond to ordered double perovskites, which are among the best-established structural families exhibiting competing magnetic interactions owing to their highly connected FCC-like magnetic sublattices~\cite{karunadasa2003ba2lnsbo6,ou2014long,yang2022magnetic}. The presence of triclinic, trigonal, orthorhombic, and tetragonal systems further illustrates that the screened materials span several crystallographic families commonly associated with frustrated magnetic behavior, including distorted perovskites and triangular magnetic networks~\cite{barik2025magnetic,weber2019topological,ramirez1994strongly,gardner2010magnetic}.

Further, SFigure\,8 compares the elemental distributions obtained using the conventional frustration criterion ($f\ge5$) and the uncertainty-aware conservative criterion ($f_{\mathrm{cons}}\ge5$). The comparison demonstrates that incorporating prediction uncertainty substantially modifies the chemical composition of the retained candidate pool. While the initial screening based solely on the conventional frustration index contains a relatively broad range of chemical systems, uncertainty-aware filtering preferentially removes many chemically isolated compounds associated with larger prediction uncertainty. Consequently, the final screened set becomes considerably more chemically focused rather than simply smaller.

Most notably, the conservative screening enriches Sr- and Mn-containing oxide systems, whereas the occurrence of Ce-, V-, Ni-, Co-, Cr-, Cu-, and several other transition-metal-containing compounds is markedly reduced. This observation indicates that many chemically diverse high-$f$ predictions are accompanied by relatively large uncertainty, while oxide perovskite-derived compounds remain consistently favored by the ensemble predictions after uncertainty correction.

Overall, the comparison between the conventional and uncertainty-aware screening demonstrates that incorporating prediction uncertainty does not simply reduce the number of retained candidates. Instead, it selectively suppresses chemically isolated predictions while preserving chemically coherent oxide families. As a result, the final high-confidence candidate pool is concentrated within a physically meaningful region of chemical space dominated by Sr-Ca-Ti-Mn oxides and ordered double perovskites, both of which are well-established structures for competing magnetic exchange interactions and magnetic frustration.

\subsection{Confirmation and literature analysis of screened candidates}
\label{l_review}
After excluding all compositions present in the datasets, which were used for model training, the uncertainty-aware screening identified $20$ high-confidence candidate materials (STable\,6). The table summarizes the predicted N\'eel and Curie-Weiss temperatures together with the conventional and conservative frustration indices. The identified compounds exhibit both positive and negative Curie-Weiss temperatures. This behavior is physically expected because the sign of $\Theta$ reflects the dominant magnetic exchange interaction. Negative values indicate prevailing AFM exchange, whereas positive values correspond to predominantly FM interactions. 

The available literature indicates that the screened compounds can be divided into three categories. The first includes experimentally established frustrated magnets, providing direct validation of the proposed workflow. The second consists of materials for which competing magnetic interactions or structural conditions favorable for frustration have previously been reported. The third category comprises compounds for which no literature discussing magnetic frustration or related magnetic properties could be identified. The strongest validation examples are provided by Ba(Cr$_5$S$_8$)$_2$, Sr$_2$YOsO$_6$, and Sr$_2$ScOsO$_6$. Ba(Cr$_5$S$_8$)$_2$ is an experimentally established geometrically frustrated AFM with a triangular Cr$^{3+}$ magnetic lattice. Magnetic susceptibility and heat-capacity measurements yielded $\Theta=-650$\,K and $T_N=63$\,K, corresponding to an experimental frustration index of $f=10.3$~\cite{yamazaki2010synthesis}. The present model predicts $\Theta=-636$\,K and $T_N=55.8$\,K, yielding $f=11.41$ and a conservative frustration index of $f_{\mathrm{cons}}=9.68$. Similarly, Sr$_2$YOsO$_6$ and Sr$_2$ScOsO$_6$ belong to the well-known family of geometrically frustrated double perovskites in which Os$^{5+}$ ions occupy an FCC magnetic sublattice. Experimental studies reported frustration indices of approximately $6.4$ and $6.6$, respectively~\cite{paul2015magnetically,taylor2015magnetic}. Although the predicted Curie-Weiss and N\'eel temperatures exhibit moderate deviations from experiment, the resulting conservative frustration indices remain in good agreement with the experimentally established frustrated character of both compounds. The quantitative comparison between available experimental data and the corresponding ML predictions is summarized in Table\,\ref{tab:comparison}.
\begin{table}[ht]
\centering
\caption{Comparison of predicted and experimental N\'eel and Curie-Weiss temperatures, frustration indices and their relative errors. All temperatures are given in K; frustration indices and relative errors are dimensionless. The $f_{\mathrm{err}}$ is calculated with respect to $f_{\mathrm{cons}}$.}
\label{tab:comparison}

\resizebox{\textwidth}{!}{%
\begin{tabular}{lcccccccccc}
\hline
Formula &
$T_{\mathrm{N}}$ &
$T_{\mathrm{N,exp}}$ &
$T_{\mathrm{N,err}}$ &
$\Theta$ &
$\Theta_{\mathrm{exp}}$ &
$\Theta_{\mathrm{err}}$ &
$f$ &
$f_{\mathrm{cons}}$ &
$f_{\mathrm{exp}}$ &
$f_{\mathrm{err}}$ \\
\hline

Ba(Cr$_5$S$_8$)$_2$ &
55.8 & 63 & 0.114 &
-636 & -650 & 0.022 &
11.41 & 9.68 & 10.3 & 0.06 \\

Sr$_2$YOsO$_6$ &
46.9 & 53 & 0.115 &
-416.8 & -345 & 0.208 &
8.88 & 6.75 & 6.4 & 0.055 \\

Sr$_2$ScOsO$_6$ &
45.5 & 53 & 0.505 &
-385 & -345 & 0.365 &
8.47 & 6.49 & 6.4 & 0.017 \\

\hline
\end{tabular}}
\end{table}

Indirect evidence supporting magnetic frustration is also available for several additional candidates. Experimental measurements on oxygen-deficient Sr$_2$CoNbO$_{6-\delta}$ revealed dominant AFM interactions without magnetic ordering down to $5$\,K, suggesting that the stoichiometric compound Sr$_2$NbCoO$_6$ is close to a frustrated magnetic regime~\cite{popov2024magnetic}. First-principles calculations further predicted that the FCC Mn sublattice in SrCaTiMnO$_6$ supports several competing AFM configurations~\cite{roy2010theory}. In Sr$_2$FeMoO$_6$, magnetic frustration has been attributed to Fe/Mo anti-site disorder, which generates competing Fe-O-Fe exchange interactions~\cite{sanchez2002effect}. Finally, BaCr$_4$S$_7$ belongs to a family of chromium sulfides with multiple competing Cr-S exchange pathways, although quantitative frustration indices have not yet been reported~\cite{narayan2016computational,bhutani2019rapid}.

Interestingly, no relevant magnetic literature could be identified for nine of the twenty screened compounds. Rather than being chemically unrelated, these compounds form several recurring chemical families. Six compounds belong to the Sr-Ca-Ti-Mn oxide system, differing only in the relative Sr/Ca/Ti/Mn ratios while preserving a common perovskite-derived structural motif. Two additional candidates, ZnCr$_8$Cu$_3$Se$_{16}$ and ZnCr$_4$CuSe$_8$, belong to the same chromium copper selenide family. The remaining compound, NaSr$_9$Fe$_5$(MoO$_6$)$_5$, represents a chemically complex Mo-containing oxide. The repeated appearance of these chemically related families indicates that the proposed workflow does not generate isolated random predictions but instead repeatedly converges toward specific compositional motifs that may constitute previously unexplored platforms for frustrated magnetism.

In summary, the literature survey demonstrates that the proposed workflow simultaneously reproduces several experimentally established frustrated magnets, identifies materials for which independent evidence of competing magnetic interactions already exists, and proposes chemically coherent families of previously unexplored compounds as promising targets for future experimental investigation.

\section{Conclusion}
\label{conclusion}
In this work, we developed an uncertainty-aware ML workflow for the indirect prediction and large-scale screening of frustrated magnetic materials based on independent prediction of the Curie-Weiss and N\'eel temperatures. Instead of directly modeling the frustration index, the proposed approach preserves the physical distinction between the two magnetic observables and explicitly propagate prediction uncertainty into the final candidate ranking.

Systematic comparison of several dataset subdivision strategies demonstrated that chemically informed partitioning significantly affects predictive performance. For the N\'eel temperature, the highest accuracy was achieved using the RE oxide subset, whereas prediction of the Curie-Weiss temperature benefited from the less restrictive RE/non-RE subdivision. Cross-class evaluation further showed that the selected models retain satisfactory predictive capability beyond their original training subsets, supporting their applicability for large-scale materials screening.

Interpretation of the trained models using SHAP analysis revealed physically meaningful descriptor importance. Ground-state magnetic moment descriptors were identified as the dominant contributors for both target properties, while descriptors related to electronic configuration, stoichiometry, and crystal chemistry provided additional nonlinear corrections. These results indicate that composition-derived descriptors capture a substantial fraction of the information governing magnetic interactions without requiring experimentally resolved crystal structures.

Application of the developed models to approximately $6.7\times10^4$ previously unseen Materials Project compounds, combined with uncertainty-aware conservative ranking, identified 20 high-confidence candidate materials predicted to exhibit frustrated magnetic behavior. The proposed conservative frustration index effectively suppresses unreliable extreme predictions while preserving chemically consistent candidate families. Importantly, the identified compounds should not be considered experimentally confirmed frustrated magnets. Instead, they represent computationally prioritized candidates for future investigation. Among them, the screening successfully recovered several experimentally established frustrated magnets, while the remaining compounds constitute either systems with indirect evidence of competing magnetic interactions or previously unexplored candidates requiring experimental validation. Oxide perovskite-derived compounds, particularly Sr-Ca-Ti-Mn systems and ordered double perovskites, emerged as especially promising targets for future studies.

Overall, the present study demonstrates that indirect prediction of magnetic frustration through separate ML models for $\Theta$ and $T_\mathrm{N}$, combined with chemically informed dataset subdivision and uncertainty-aware ranking, provides a practical and physically interpretable framework for accelerated discovery of frustrated magnetic materials. The proposed methodology is readily extendable to larger experimental databases and can be integrated with first-principles calculations and experimental validation for future exploration of quantum magnetic materials.

\section*{Acknowledgments}

The work of I.I. Piyanzina (Gumarova) was supported by the grant 24PostDoc/2-2F006 and the research project 23LCG-2F012 by the Higher Education and Science Committee of Armenia.

\bibliographystyle{elsarticle-num} 
\bibliography{refs}

@article{itani2025northeast,
  title={The northeast materials database for magnetic materials},
  author={Itani, Suman and Zhang, Yibo and Zang, Jiadong},
  journal={Nature Communications},
  volume={16},
  number={1},
  pages={9415},
  year={2025},
  publisher={Nature Publishing Group UK London}
}

@article{prokhorenkova2018catboost,
  title={{CatBoost}: unbiased boosting with categorical features},
  author={Prokhorenkova, Liudmila and Gusev, Gleb and Vorobev, Aleksandr and Dorogush, Anna Veronika and Gulin, Andrey},
  journal={Advances in neural information processing systems},
  volume={31},
  year={2018}
}

@article{veronika2018catboost,
  title={{CatBoost}: gradient boosting with categorical features support},
  author={Veronika Dorogush, Anna and Ershov, Vasily and Gulin, Andrey},
  journal={arXiv e-prints},
  pages={arXiv--1810},
  year={2018}
}

@article{ong2013python,
  title={Python {M}aterials {G}enomics (pymatgen): A robust, open-source python library for materials analysis},
  author={Ong, Shyue Ping and Richards, William Davidson and Jain, Anubhav and Hautier, Geoffroy and Kocher, Michael and Cholia, Shreyas and Gunter, Dan and Chevrier, Vincent L and Persson, Kristin A and Ceder, Gerbrand},
  journal={Computational Materials Science},
  volume={68},
  pages={314--319},
  year={2013},
  publisher={Elsevier}
}

@article{ward2018matminer,
  title={Matminer: An open source toolkit for materials data mining},
  author={Ward, Logan and Dunn, Alexander and Faghaninia, Alireza and Zimmermann, Nils ER and Bajaj, Saurabh and Wang, Qi and Montoya, Joseph and Chen, Jiming and Bystrom, Kyle and Dylla, Maxwell and others},
  journal={Computational Materials Science},
  volume={152},
  pages={60--69},
  year={2018},
  publisher={Elsevier}
}

@article{ward2016general,
  title={A general-purpose machine learning framework for predicting properties of inorganic materials},
  author={Ward, Logan and Agrawal, Ankit and Choudhary, Alok and Wolverton, Christopher},
  journal={npj Computational Materials},
  volume={2},
  number={1},
  pages={16028},
  year={2016},
  publisher={Nature Publishing Group}
}

@article{horton2025accelerated,
  title={Accelerated data-driven materials science with the {M}aterials {P}roject},
  author={Horton, Matthew K and Huck, Patrick and Yang, Ruo Xi and Munro, Jason M and Dwaraknath, Shyam and Ganose, Alex M and Kingsbury, Ryan S and Wen, Mingjian and Shen, Jimmy X and Mathis, Tyler S and others},
  journal={Nature Materials},
  volume={24},
  number={10},
  pages={1522--1532},
  year={2025},
  publisher={Nature Publishing Group UK London}
}

@article{jain2013commentary,
  title={Commentary: The {M}aterials {P}roject: A materials genome approach to accelerating materials innovation},
  author={Jain, Anubhav and Ong, Shyue Ping and Hautier, Geoffroy and Chen, Wei and Richards, William Davidson and Dacek, Stephen and Cholia, Shreyas and Gunter, Dan and Skinner, David and Ceder, Gerbrand and others},
  journal={APL materials},
  volume={1},
  number={1},
  year={2013},
  publisher={AIP Publishing}
}

@misc{wiebe2024fantastic,
  title={Fantastic Frustrated Materials--and Where to Find Them},
  author={Wiebe, Christopher R},
  year={2024},
  publisher={ACS Publications}
}

@article{leszczynski2024machine,
  title={Machine-learned models for power magnetic material characteristics},
  author={Leszczy{\'n}ski, Pawe{\l} and Kutorasi{\'n}ski, Kamil and Szewczyk, Marcin and Paw{\l}owski, Jaros{\l}aw},
  journal={IEEE Transactions on Power Electronics},
  volume={40},
  number={1},
  pages={1554--1562},
  year={2024},
  publisher={IEEE}
}

@article{nelson2019predicting,
  title={Predicting the {C}urie temperature of ferromagnets using machine learning},
  author={Nelson, James and Sanvito, Stefano},
  journal={Physical Review Materials},
  volume={3},
  number={10},
  pages={104405},
  year={2019},
  publisher={APS}
}

@article{belot2023machine,
  title={Machine learning predictions of high-{C}urie-temperature materials},
  author={Belot, Joshua F and Taufour, Valentin and Sanvito, Stefano and Hart, Gus LW},
  journal={Applied Physics Letters},
  volume={123},
  number={4},
  year={2023},
  publisher={AIP Publishing}
}

@article{long2021accelerating,
  title={An accelerating approach of designing ferromagnetic materials via machine learning modeling of magnetic ground state and Curie temperature},
  author={Long, Teng and Fortunato, Nuno M and Zhang, Yixuan and Gutfleisch, Oliver and Zhang, Hongbin},
  journal={Materials Research Letters},
  volume={9},
  number={4},
  pages={169--174},
  year={2021},
  publisher={Taylor \& Francis}
}

@article{jung2024machine,
  title={Machine-learning prediction of {C}urie temperature from chemical compositions of ferromagnetic materials},
  author={Jung, Son Gyo and Jung, Guwon and Cole, Jacqueline M},
  journal={Journal of Chemical Information and Modeling},
  volume={64},
  number={16},
  pages={6388--6409},
  year={2024},
  publisher={ACS Publications}
}

@article{lundberg2017unified,
  title={A unified approach to interpreting model predictions},
  author={Lundberg, Scott M and Lee, Su-In},
  journal={Advances in neural information processing systems},
  volume={30},
  year={2017}
}

@article{orang2026predicting,
  title={Predicting the {C}urie temperature of magnetic materials with machine learning: Descriptor engineering, graph neural networks, and the role of curated data},
  author={Orang, Akram Abedi and Alaei, Mojtaba and Oganov, Artem R},
  journal={Computational Materials Science},
  volume={269},
  pages={114663},
  year={2026},
  publisher={Elsevier}
}

@article{orang2026curie,
  title={Predicting the {C}urie temperature of magnetic materials with machine learning: Descriptor engineering, graph neural networks, and the role of curated data},
  author={Orang, Akram Abedi and Alaei, Mojtaba and Oganov, Artem R},
  journal={Computational Materials Science},
  volume={269},
  pages={114663},
  year={2026},
  publisher={Elsevier}
}

@article{ramirez1994strongly,
title={Strongly geometrically frustrated magnets},
author={Ramirez, Arthur P.},
journal={Annual Review of Materials Science},
volume={24},
number={1},
pages={453--480},
year={1994},
doi={10.1146/annurev.ms.24.080194.002321}
}

@article{gardner2010magnetic,
title={Magnetic pyrochlore oxides},
author={Gardner, Jason S. and Gingras, Michel J. P. and Greedan, John E.},
journal={Reviews of Modern Physics},
volume={82},
number={1},
pages={53--107},
year={2010},
doi={10.1103/RevModPhys.82.53}
}

@article{weber2019topological,
  title={Topological semimetal features in the multiferroic hexagonal manganites},
  author={Weber, Sophie F and Griffin, Sin{\'e}ad M and Neaton, Jeffrey B},
  journal={Physical Review Materials},
  volume={3},
  number={6},
  pages={064206},
  year={2019},
  publisher={APS}
}

@book{taylor1997introduction,
  title = {An Introduction to Error Analysis: The Study of Uncertainties in Physical Measurements},
  author = {Taylor, John R.},
  publisher = {University Science Books},
  edition = {2},
  year = {1997}
}

@article{kanamori2006rare,
  title={Rare earth elements and magnetism in metallic systems},
  author={Kanamori, Junjiro},
  journal={Journal of alloys and compounds},
  volume={408},
  pages={2--8},
  year={2006},
  publisher={Elsevier}
}

@article{nqayi2025super,
  title={Super-exchange coupling in oxygen rich rare-earth based Sm2MnRuO6+ $\delta$ double perovskite},
  author={Nqayi, S and Sondezi, B},
  journal={Materials Chemistry and Physics},
  volume={332},
  pages={130294},
  year={2025},
  publisher={Elsevier}
}

@article{tatarchuk2024studying,
  title={Studying the defects in spinel compounds: discovery, formation mechanisms, classification, and influence on catalytic properties},
  author={Tatarchuk, Tetiana},
  journal={Nanomaterials},
  volume={14},
  number={20},
  pages={1640},
  year={2024},
  publisher={MDPI}
}

@article{kruskal1952use,
  title={Use of ranks in one-criterion variance analysis},
  author={Kruskal, William H and Wallis, W Allen},
  journal={Journal of the American statistical Association},
  volume={47},
  number={260},
  pages={583--621},
  year={1952},
  publisher={Taylor \& Francis}
}

@article{mann1947test,
  title={On a test of whether one of two random variables is stochastically larger than the other},
  author={Mann, Henry B and Whitney, Donald R},
  journal={The annals of mathematical statistics},
  pages={50--60},
  year={1947},
  publisher={JSTOR}
}

@article{lazar2026beyond,
  title={Beyond Accuracy: A Comprehensive Comparative Study of Gradient Boosting Versus Tabular Deep Learning and Explainability Techniques for Mixed-Type Tabular Data Models Using SHAP and LIME},
  author={Lazar, Alina and Pokhrel, Prativa and Das, Shrabony},
  journal={International Journal on Artificial Intelligence Tools},
  volume={35},
  number={02},
  pages={2640003},
  year={2026},
  publisher={World Scientific}
}

@article{singh2023physics,
  title={Physics-informed machine-learning prediction of curie temperatures and its promise for guiding the discovery of functional magnetic materials},
  author={Singh, Prashant and Del Rose, Tyler and Palasyuk, Andriy and Mudryk, Yaroslav},
  journal={Chemistry of Materials},
  volume={35},
  number={16},
  pages={6304--6312},
  year={2023},
  publisher={ACS Publications}
}

@article{de2021conservatism,
  title={A conservatism index based on structural reliability and model errors},
  author={de Santana Gomes, Wellison Jos{\'e} and Beck, Andr{\'e} Te{\'o}filo},
  journal={Reliability Engineering \& System Safety},
  volume={209},
  pages={107456},
  year={2021},
  publisher={Elsevier}
}

@article{karunadasa2003ba2lnsbo6,
  title={Ba2LnSbO6 and Sr2LnSbO6 (Ln= Dy, Ho, Gd) double perovskites: Lanthanides in the geometrically frustrating fcc lattice},
  author={Karunadasa, H and Huang, Q and Ueland, BG and Schiffer, P and Cava, RJ},
  journal={Proceedings of the National Academy of Sciences},
  volume={100},
  number={14},
  pages={8097--8102},
  year={2003},
  publisher={National Academy of Sciences}
}

@article{ou2014long,
  title={Long-range magnetic interaction and frustration in double perovskites Sr2NiIrO6 and Sr2ZnIrO6},
  author={Ou, Xuedong and Li, Zhengwei and Fan, Fengren and Wang, Hongbo and Wu, Hua},
  journal={Scientific Reports},
  volume={4},
  number={1},
  pages={7542},
  year={2014},
  publisher={Nature Publishing Group UK London}
}

@article{yang2022magnetic,
  title={Magnetic frustration in the cubic double perovskite Ba 2 NiIrO 6},
  author={Yang, Ke and Xu, Wenjing and Lu, Di and Zhou, Yuxuan and Liu, Lu and Ma, Yaozhenghang and Wang, Guangyu and Wu, Hua},
  journal={Physical Review B},
  volume={105},
  number={18},
  pages={184413},
  year={2022},
  publisher={APS}
}

@article{goto2021geometrical,
  title={Geometrical spin frustration and monoclinic-distortion-induced spin canting in the double perovskites Ln2LiFeO6 (Ln= La, Nd, Sm, and Eu) with unusually high valence Fe5+},
  author={Goto, Masato and Oguchi, Tamio and Shimakawa, Yuichi},
  journal={Journal of the American Chemical Society},
  volume={143},
  number={45},
  pages={19207--19213},
  year={2021},
  publisher={ACS Publications}
}

@article{duffort2013lifting,
  title={Lifting the geometric frustration through a monoclinic distortion in “114” YBaFe4O7. 0: Magnetism and transport},
  author={Duffort, V and Sarkar, T and Caignaert, V and Pralong, V and Raveau, B and Avdeev, M and Cervellino, A and Waerenborgh, JC and Tsipis, EV},
  journal={Journal of Solid State Chemistry},
  volume={205},
  pages={225--235},
  year={2013},
  publisher={Elsevier}
}

@article{barik2025magnetic,
  title={Magnetic ground state of a Jeff= 1/2 based frustrated triangular lattice antiferromagnet},
  author={Barik, M and Khatua, J and Kim, Suyoung and Mun, Eundeok and Lee, Suheon and Hitti, Bassam and Morris, Gerald D and Choi, Kwang-Yong and Khuntia, P},
  journal={arXiv preprint arXiv:2504.07194},
  year={2025}
}

@article{popov2024magnetic,
  title={Magnetic properties of the double perovskite Sr 2 CoNbO 6-$\delta$},
  author={Popov, DV and Batulin, RG and Cherosov, MA and Yatsyk, IV and Chupakhina, TI and Deeva, Yu A and Eremina, RM and Maiti, T},
  journal={Magnetic Resonance in Solids},
  volume={25},
  number={3},
  pages={23301--9},
  year={2024}
}

@article{roy2010theory,
  title={Theory of prospective tetrahedral perovskite ferroelectrics},
  author={Roy, Anindya and Vanderbilt, David},
  journal={arXiv preprint arXiv:1012.5070},
  year={2010}
}

@article{sanchez2002effect,
  title={The effect of substitutional mis-site disorder on the structural, magnetic and electronic transport properties of Sr2FeMoO6 double perovskites},
  author = {S\'anchez, D and Garc\'ia-Hern\'andez, M
            and Mart\'inez, J.L. and Alonso, J.A.
            and Mart\'inez-Lope, M.J.
            and Casais, M.T.
            and Mellerg{\aa}rd, A},
  journal={Journal of magnetism and magnetic materials},
  volume={242},
  pages={729--731},
  year={2002},
  publisher={Elsevier}
}

@article{yamazaki2010synthesis,
  title={Synthesis, structures and magnetic properties of pseudo-hollandite chromium sulfides},
  author={Yamazaki, Satoshi and Ueda, Yutaka},
  journal={Journal of Solid State Chemistry},
  volume={183},
  number={9},
  pages={1905--1911},
  year={2010},
  publisher={Elsevier}
}

@article{narayan2016computational,
  title={Computational and experimental investigation for new transition metal selenides and sulfides: The importance of experimental verification for stability},
  author={Narayan, Awadhesh and Bhutani, Ankita and Rubeck, Samantha and Eckstein, James N and Shoemaker, Daniel P and Wagner, Lucas K},
  journal={Physical Review B},
  volume={94},
  number={4},
  pages={045105},
  year={2016},
  publisher={APS}
}

@phdthesis{bhutani2019rapid,
  title={Rapid materials discovery of ternary transition metal chalcogenides},
  author={Bhutani, Ankita},
  year={2019},
  school={University of Illinois at Urbana-Champaign}
}

@article{paul2015magnetically,
  title={Magnetically frustrated double perovskites: synthesis, structural properties, and magnetic order of Sr2BOsO6 (B= Y, In, Sc)},
  author={Paul, Avijit Kumar and Sarapulova, Angelina and Adler, Peter and Reehuis, Manfred and Kanungo, Sudipta and Mikhailova, Daria and Schnelle, Walter and Hu, Zhiwei and Kuo, Changyang and Siruguri, Vasudeva and others},
  journal={Zeitschrift f{\"u}r anorganische und allgemeine Chemie},
  volume={641},
  number={2},
  pages={197--205},
  year={2015},
  publisher={Wiley Online Library}
}

@article{taylor2015magnetic,
  title={Magnetic order and electronic structure of the 5 d 3 double perovskite Sr 2 ScOsO 6},
  author={Taylor, AE and Morrow, R and Singh, DJ and Calder, S and Lumsden, MD and Woodward, PM and Christianson, AD},
  journal={Physical Review B},
  volume={91},
  number={10},
  pages={100406},
  year={2015},
  publisher={APS}
}

@article{navratil2025interpretable,
  title={Interpretable machine learning for atomic scale magnetic anisotropy in quantum materials},
  author={Navr{\'a}til, Jan and Topolnicki, Rafa{\l} and Otyepka, Michal and B{\l}o{\'n}ski, Piotr},
  journal={npj Computational Materials},
  volume={11},
  number={1},
  pages={138},
  year={2025},
  publisher={Nature Publishing Group UK London}
}

@article{berzin2012switching,
  title={“Switching” of the nanodomain state of the frustrated ferromagnet-antiferromagnet system by an external magnetic field},
  author={Berzin, Alexey Adolfovich and Morosov, AI and Sigov, Aleksandr Sergeevich},
  journal={Physics of the Solid State},
  volume={54},
  number={11},
  pages={2226--2230},
  year={2012},
  publisher={Springer}
}

@incollection{sigov2023behavior,
  title={Behavior in a Magnetic Field},
  author={Sigov, Alexander S},
  booktitle={Multilayer Magnetic Nanostructures: Properties and Applications},
  pages={61--79},
  year={2023},
  publisher={Springer}
}

@incollection{sigov2023spin,
  title={Spin-Valve Structure Ferromagnet--Antiferromagnet--Ferromagnet},
  author={Sigov, Alexander S},
  booktitle={Multilayer Magnetic Nanostructures: Properties and Applications},
  pages={81--108},
  year={2023},
  publisher={Springer}
}

@incollection{sigov2023physical,
  title={Physical Foundations for the Formation of Magnetic Nanostructures},
  author={Sigov, Alexander S},
  booktitle={Multilayer Magnetic Nanostructures: Properties and Applications},
  pages={1--18},
  year={2023},
  publisher={Springer}
}

@article{desmarais2025meta,
  title={Meta-generalized gradient approximation made magnetic},
  author={Desmarais, Jacques K and Erba, Alessandro and Vignale, Giovanni and Pittalis, Stefano},
  journal={Physical Review Letters},
  volume={134},
  number={10},
  pages={106402},
  year={2025},
  publisher={APS}
}

@article{garrido2025challenges,
  title={Challenges and strategies for first-principles simulations of two-dimensional magnetic phenomena},
  author={Garrido Aldea, Jaime and Esteras, Dorye L and Roche, Stephan and Garcia, Jose H},
  journal={Nanoscale},
  volume={17},
  number={43},
  pages={24955--24989},
  year={2025},
  publisher={The Royal Society of Chemistry}
}

@article{zhang2026refrigeration,
  title={Refrigeration down to 0.16 K using a frustrated magnet Gd2B2MoO9},
  author={Zhang, Yikun and Na, Yingzhe and Liu, Xinyang and Xiang, Junsen and Chen, Fengying and Li, Hai-Feng and Sun, Peijie and Zhou, Shengqiang and Zhang, Xuefeng and Li, Lingwei},
  journal={Nature Communications},
  volume={17},
  number={1},
  pages={1554},
  year={2026},
  publisher={Nature Publishing Group UK London}
}

@article{zhang2024geometrically,
  title={Geometrically frustrated Gd2Ti2O7 oxide: A comprehensive exploration of structural, magnetic, and magnetocaloric properties for cryogenic magnetic cooling applications},
  author={Zhang, Yikun and Hao, Weixiang and Lin, Junli and Li, Hai-Feng and Li, Lingwei},
  journal={Acta Materialia},
  volume={272},
  pages={119946},
  year={2024},
  publisher={Elsevier}
}

@article{zhang2026geometrically,
  title={A Geometrically Frustrated LiCaGd5 (BO3) 6 Oxide with a Record Low-Field Magnetocaloric Response at Ultra-Low Temperature},
  author={Zhang, Zhenqian and Sun, Guangyi and Jiang, Kaifeng and Zhang, Zhenhua and Li, Lingwei},
  journal={Advanced Functional Materials},
  pages={e76221},
  year={2026},
  publisher={Wiley Online Library}
}

\end{document}

\begin{frontmatter}

\title{Supplementary Materials:\\
Indirect prediction and screening of frustrated magnetic materials using machine learning}

\author{Regina Burganova$^{a*}$,  Irina Piyanzina $^{b}$, Ruslan Akhmerov$^{c}$, Alexander Sigov$^{d}$, 
Maxim Fedorov$^{a}$  \\ 
bur.regina@gmail.com}

\affiliation{organization={$^a$Institute for Information Transmission Problems (Kharkevich Institute)},%Department and Organization
           addressline={Bolshoy Karetny per. 19, build.1}, 
            city={Moscow},
            postcode={127051}, 
%           state={},
           country={Russia}} 
           
\affiliation{organization={$^b$Center of Semiconductor Devices and Nanotechnology, Computational Materials Science Laboratory, Yerevan State University},    
addressline={1 Alex Manoogian St.}, 
           city={Yerevan},
            postcode={0025}, 
      %     state={},
           country={Republic of Armenia}}

           \affiliation{organization={$^c$Progsol LLC},
            addressline={Noyabrskaya, 52~str.},  
            city={Orenburg},
            postcode={460050}, 
            country={Russia}}

\affiliation{organization={$^d$Department of Nanoelectronics, MIREA—Russian Technological University},%Department and Organization
           addressline={78 Vernadskiy Avenue}, 
            city={Moscow},
            postcode={119454}, 
%           state={},
           country={Russia}} 

\end{frontmatter}

\begin{figure}[H]
\label{fig:first}
    \centering
    \begin{minipage}{0.8\textwidth}
        \centering
        \includegraphics[width=\linewidth]{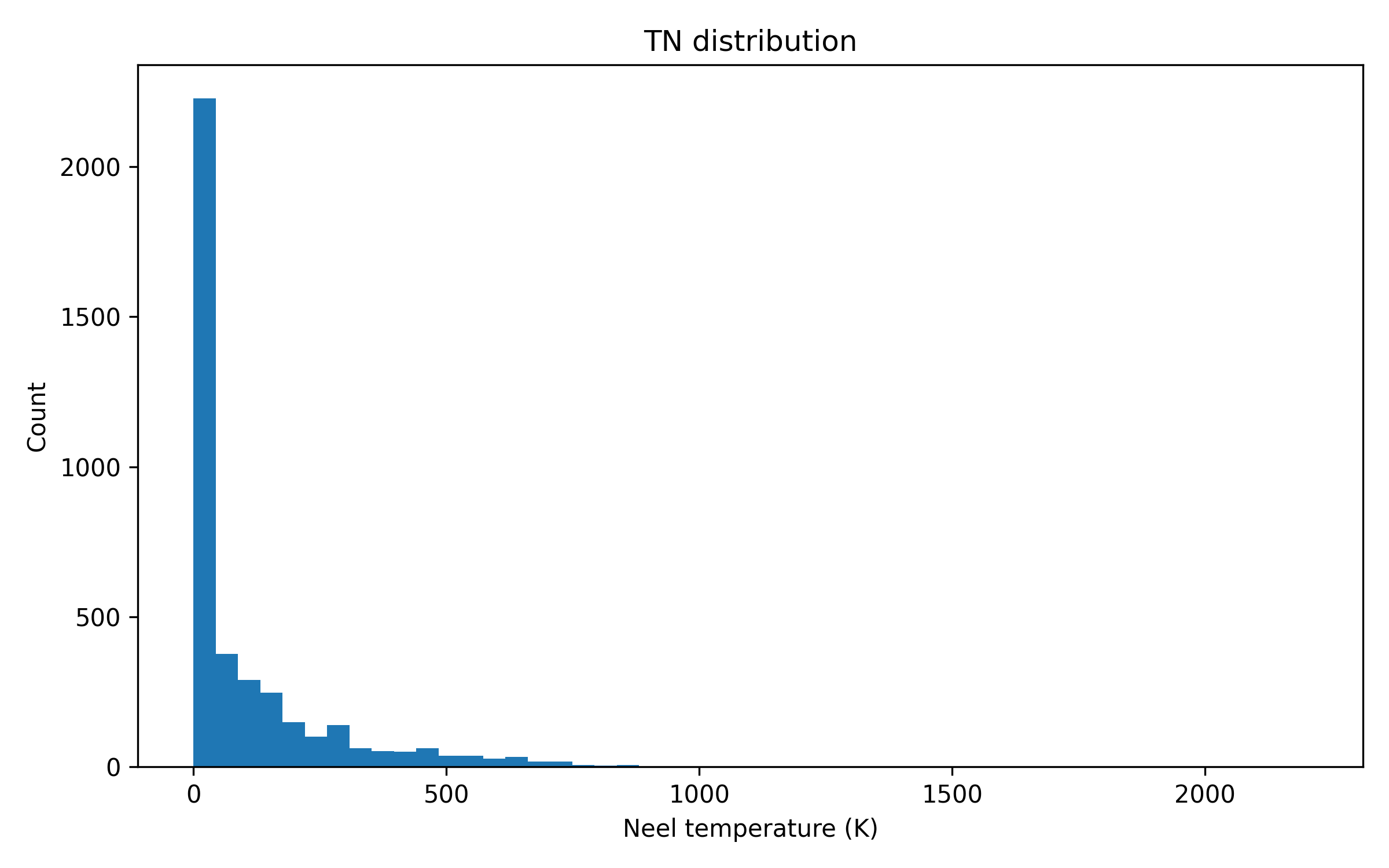} \\
        (a)
           \end{minipage} 
    
    \begin{minipage}{0.8\textwidth}
        \centering
        \includegraphics[width=\linewidth]{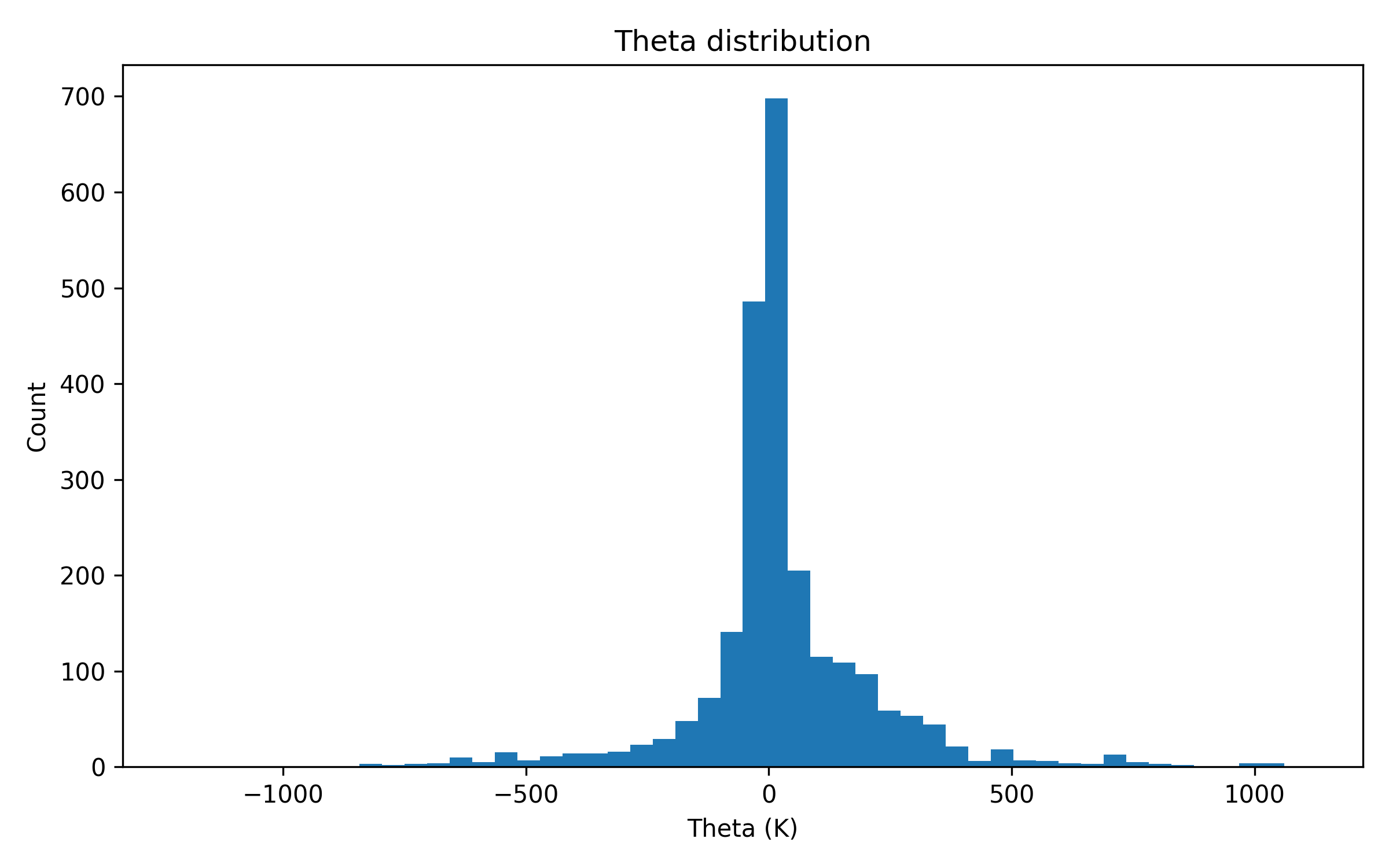} \\
        (b)
          \end{minipage}
    \caption{Distribution of clean data for the (a) N\'eel and (b) Curie-Weiss temperatures.}
\end{figure}

\begin{figure}[H]
\label{fig:groups}
    \centering
    \begin{minipage}{0.8\textwidth}
        \centering
        \includegraphics[width=\linewidth]{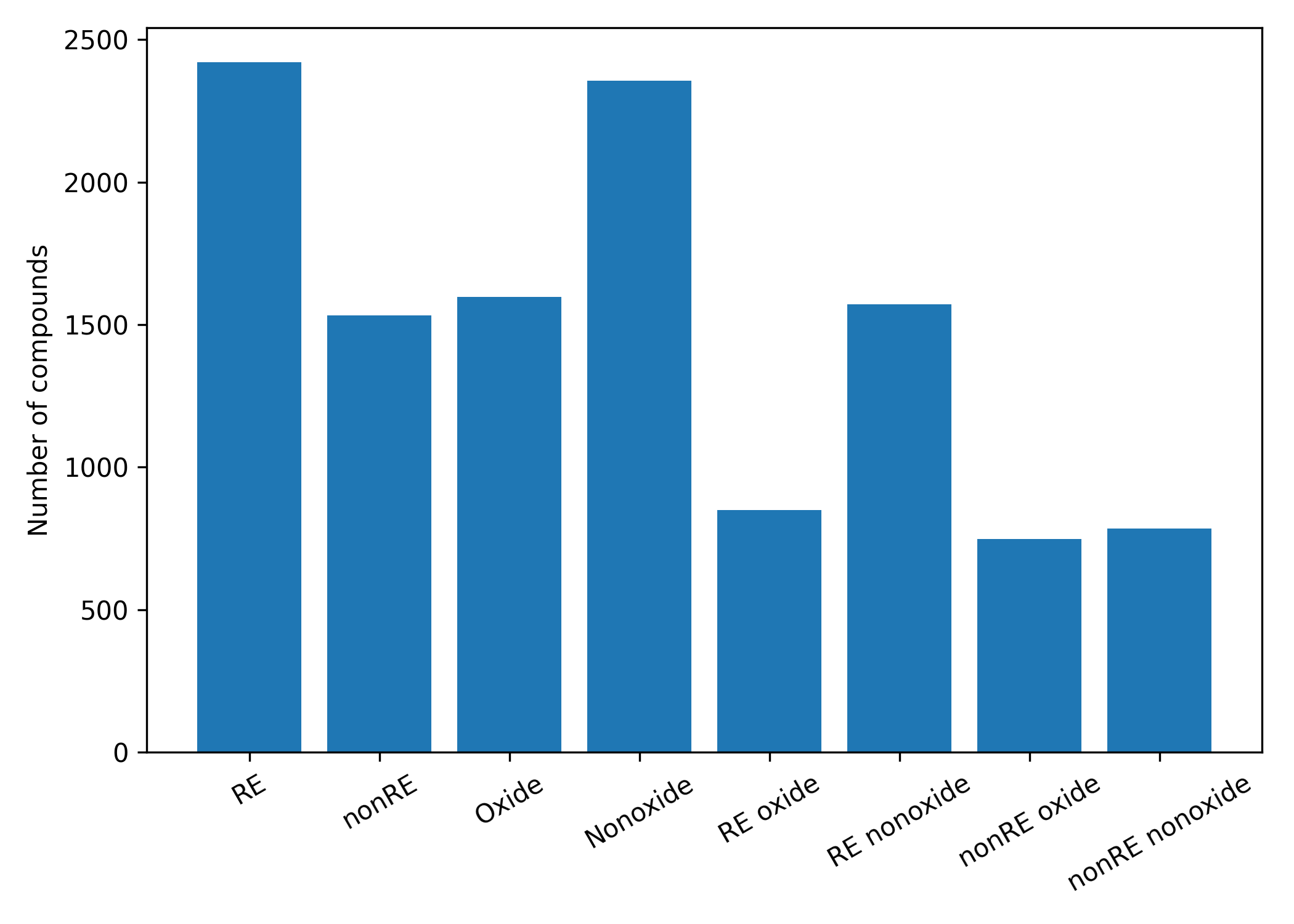} \\
        (a)
           \end{minipage}
    \hfill 
     \begin{minipage}{0.8\textwidth}
        \centering
        \includegraphics[width=\linewidth]{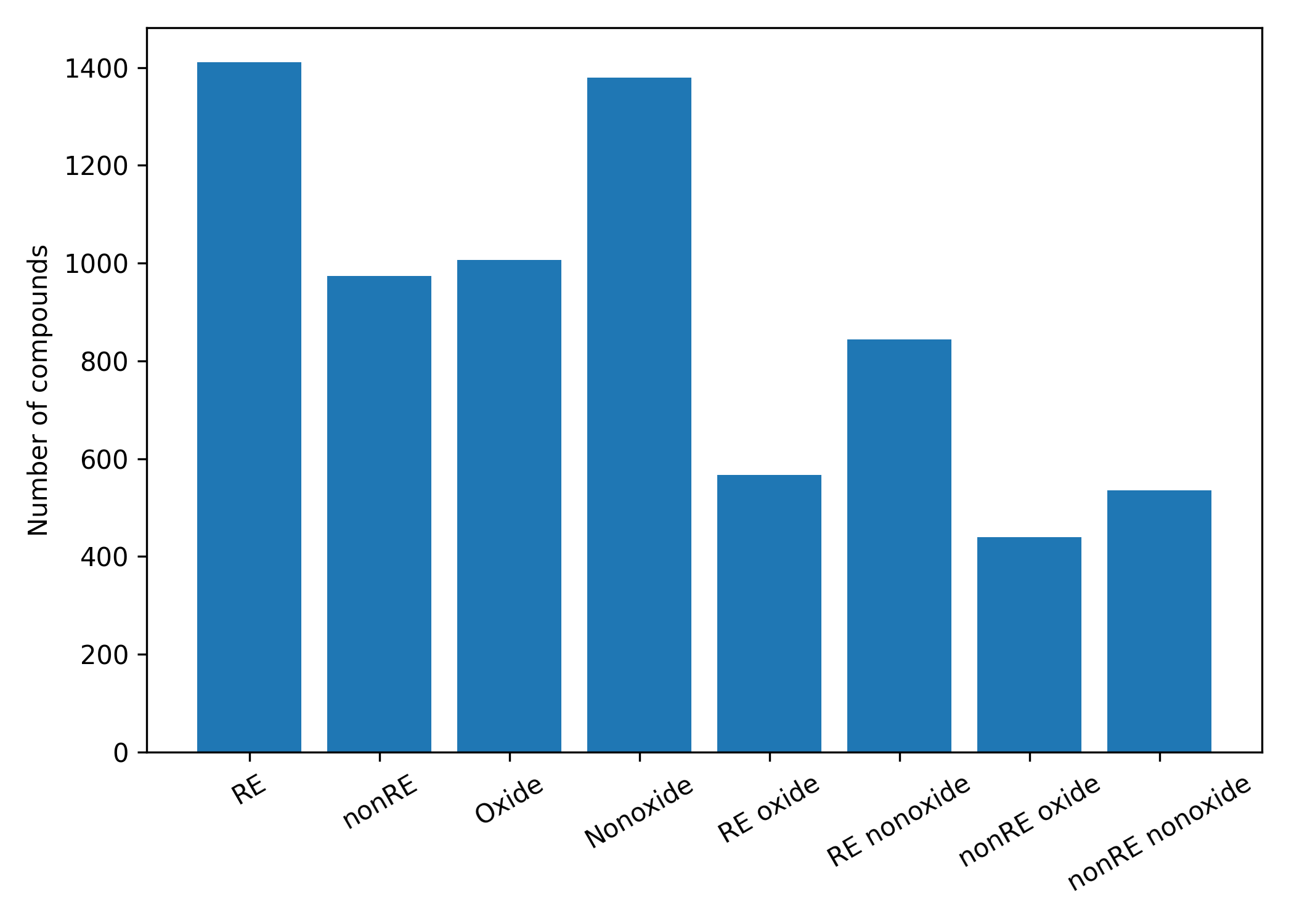} \\
        (b)
           \end{minipage}
        \caption{Statistic of groups for (a) N\'eel temperature and (b) Curie–Weiss temperature.}
\end{figure}

\begin{figure}[htbp]
\centering
\includegraphics[width=\linewidth]{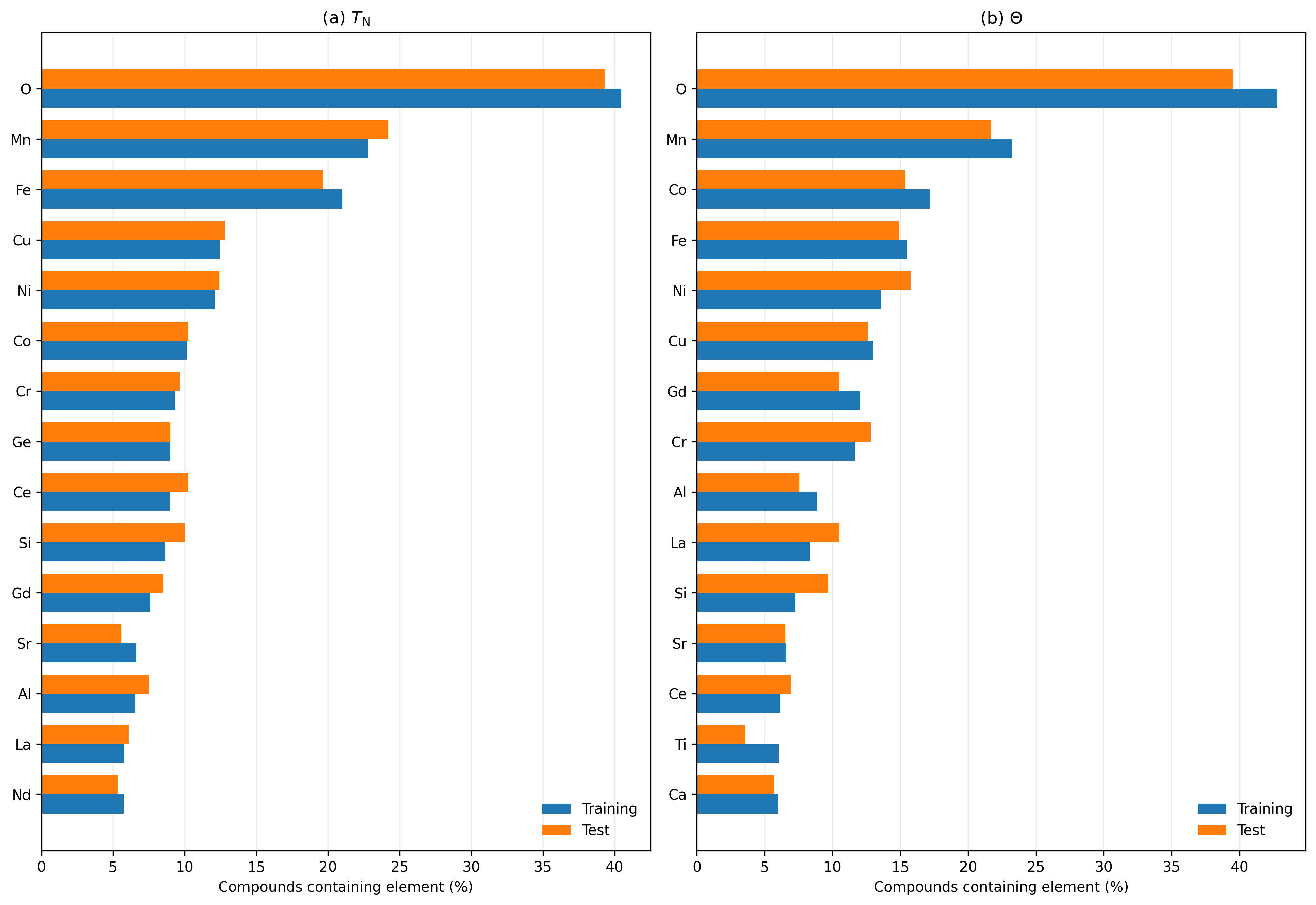}
\caption{Comparison of elemental occurrence frequencies in the training and independent holdout test subsets for (a) $T_\mathrm{N}$ and (b) $\Theta$. Frequencies are expressed as the percentage of compounds containing each element. The 15 most frequent elements in the training subset are shown for each dataset.}
\label{fig:element_distribution_train_test}
\end{figure} 

\begin{table}[H]
\caption{Prediction performance for the complete dataset and four-class subdivision.}
\centering
\scalebox{0.82}{
\begin{tabular}{lllccccc}
\hline
Target & Evaluation & Metric &
Full data & RE oxide & nonRE oxide & RE nonoxide & nonRE nonoxide \\
\hline

\multirow{9}{*}{$T_\mathrm{N}$}
& \multirow{3}{*}{CV}
& MAE  & 47.71 & 47.05 & 70.14 & 21.17 & 71.00 \\
&
& RMSE & 96.36 & 77.97 &126.10 & 47.32 &124.40 \\
&
& $R^2$ & 0.67 & 0.78 & 0.65 & 0.78 & 0.51 \\
\cline{2-8}

& \multirow{3}{*}{OOF}
& MAE  & 47.71 & 47.06 & 70.13 & 21.17 & 70.98 \\
&
& RMSE & 98.00 & 79.32 &138.55 & 48.36 &125.68 \\
&
& $R^2$ & 0.67 & 0.79 & 0.61 & 0.79 & 0.53 \\
\cline{2-8}

& \multirow{3}{*}{Test}
& MAE  & 43.00 & 38.51 & 67.01 & 22.50 & 81.38 \\
&
& RMSE & 88.07 & 67.71 &100.84 & 57.81 &186.01 \\
&
& $R^2$ & 0.73 & 0.83 & 0.77 & 0.64 & 0.26 \\
\hline

\multirow{9}{*}{$\Theta$}
& \multirow{3}{*}{CV}
& MAE  & 65.04 & 57.67 & 93.65 & 30.86 & 89.33 \\
&
& RMSE &123.38 &106.11 &155.65 & 63.10 &145.11 \\
&
& $R^2$ & 0.60 & 0.67 & 0.59 & 0.61 & 0.61 \\
\cline{2-8}

& \multirow{3}{*}{OOF}
& MAE  & 65.04 & 57.67 & 93.65 & 30.84 & 89.38 \\
&
& RMSE &126.63 &113.54 &166.27 & 67.71 &153.26 \\
&
& $R^2$ & 0.60 & 0.66 & 0.56 & 0.61 & 0.61 \\
\cline{2-8}

& \multirow{3}{*}{Test}
& MAE  & 63.09 & 56.82 &102.39 & 27.55 &111.28 \\
&
& RMSE &127.85 &130.36 &150.31 & 58.51 &229.22 \\
&
& $R^2$ & 0.63 & 0.51 & 0.51 & 0.59 & 0.51 \\
\hline

\end{tabular}}
\label{tab:performance}
\end{table}

\begin{figure}[H]
\centering
\begin{minipage}{0.48\textwidth}
\centering
\includegraphics[width=\linewidth]{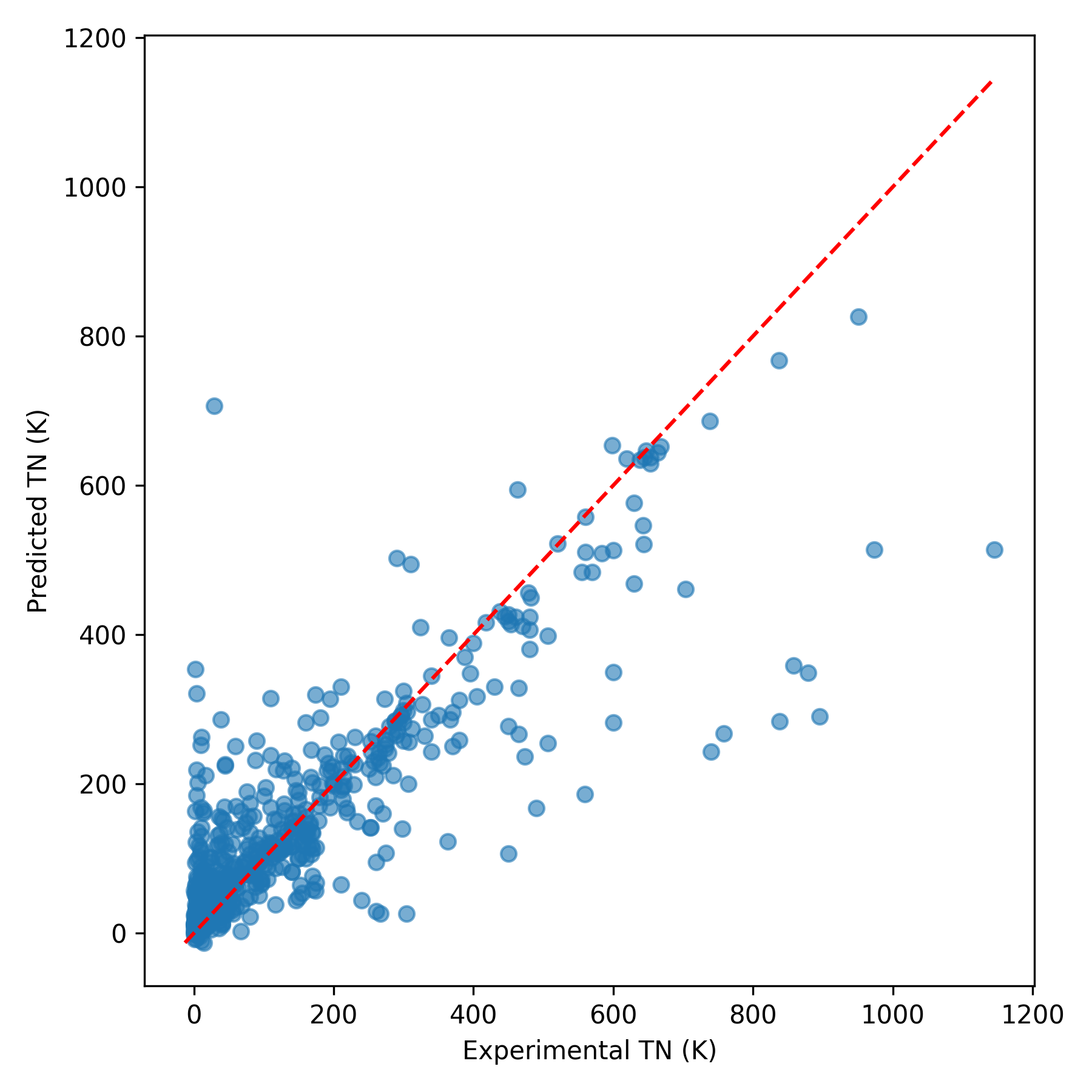}\\
(a)
\end{minipage}
\hfill
\begin{minipage}{0.48\textwidth}
\centering
\includegraphics[width=\linewidth]{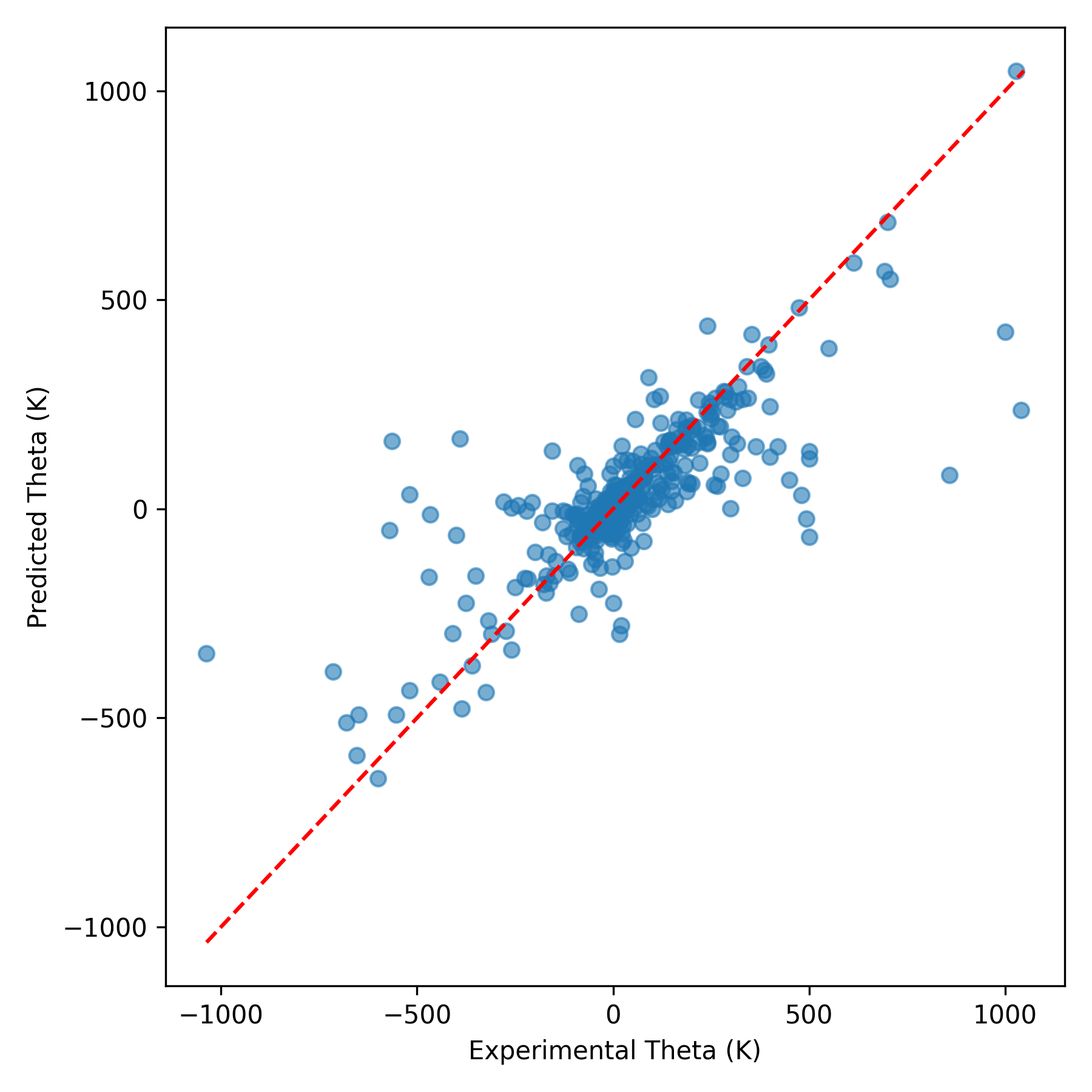}\\
(b)
\end{minipage}
\caption{Parity plots obtained for prediction of (a) the N\'eel temperature and (b) the Curie-Weiss temperature for full materials dataset.}
\label{fig:parity}
\end{figure}

\begin{table}[H]
\centering
\caption{Descriptive statistics of N\'eel ($T_\mathrm{N}$) and Curie-Weiss ($\Theta$) temperatures for chemically defined subsets.}
\label{tab:group_statistics}
\small
\begin{tabular}{llcccccc}
\hline
Property & Group & $N$ & Mean & Median & Q1 & Q3 & Range \\
\hline

$T_\mathrm{N}$ &
RE oxide        & 849 & 137.00 & 84.40 & 12.00 & 170.0 & $[0.0045,\;865]$ \\
&
RE nonoxide     & 1572 & 51.83 & 11.55 & 5.00 & 34.7 & $[0.0025,\;758]$ \\
&
nonRE oxide     & 748 & 157.66 & 51.50 & 18.91 & 220.0 & $[0.38,\;2202]$ \\
&
nonRE nonoxide  & 785 & 156.48 & 85.00 & 28.00 & 235.0 & $[0.217,\;2054]$ \\

\hline

$\Theta$ &
RE oxide        & 567 & 34.60 & -0.58 & -19.56 & 131.13 & $[-1027,\;857.5]$ \\
&
RE nonoxide     & 844 & 23.04 & 5.20 & -15.03 & 39.10 & $[-716,\;800]$ \\
&
nonRE oxide     & 439 & -33.16 & -20.00 & -93.85 & 30.00 & $[-1037,\;1000]$ \\
&
nonRE nonoxide  & 535 & 72.74 & 44.50 & -17.75 & 178.80 & $[-1214,\;1107]$ \\

\hline
\end{tabular}
\end{table}

\begin{table}[H]
\centering
\caption{Results of the Kruskal-Wallis test comparing chemically defined subsets.}
\label{tab:kruskal}
\small
\begin{tabular}{lcc}
\hline
Property & H statistic & $p$ value \\
\hline
$\Theta$         & 131.056 & $3.20 \times 10^{-28}$ \\
$T_\mathrm{N}$   & 643.232 & $4.27 \times 10^{-139}$ \\
\hline
\end{tabular}
\end{table}

\begin{table}[H]
\centering
\caption{Pairwise Mann-Whitney U test between chemically defined subsets.}
\label{tab:mannwhitney}
\small
\begin{tabular}{llccccc}
\hline
Property & Comparison & $N_1$ & $N_2$ & U statistic & $p$ value \\
\hline

$\Theta$
& RE oxide -- RE nonoxide        & 567 & 844 & 240375.5 & $8.83 \times 10^{-1}$ \\
& RE oxide -- nonRE oxide        & 567 & 439 & 157760.5 & $3.17 \times 10^{-13}$ \\
& RE oxide -- nonRE nonoxide     & 567 & 535 & 128760.0 & $1.43 \times 10^{-5}$ \\
& RE nonoxide -- nonRE oxide     & 844 & 439 & 239753.0 & $4.94 \times 10^{-18}$ \\
& RE nonoxide -- nonRE nonoxide  & 844 & 535 & 179566.0 & $1.44 \times 10^{-10}$ \\
& nonRE oxide -- nonRE nonoxide  & 439 & 535 & 75037.0  & $2.87 \times 10^{-22}$ \\

\hline

$T_\mathrm{N}$
& RE oxide -- RE nonoxide        & 849 & 1572 & 924820.5 & $1.77 \times 10^{-55}$ \\
& RE oxide -- nonRE oxide        & 849 & 748  & 306668.0 & $2.38 \times 10^{-1}$ \\
& RE oxide -- nonRE nonoxide     & 849 & 785  & 298831.0 & $3.06 \times 10^{-4}$ \\
& RE nonoxide -- nonRE oxide     & 1572 & 748 & 308215.0 & $8.45 \times 10^{-77}$ \\
& RE nonoxide -- nonRE nonoxide  & 1572 & 785 & 277724.5 & $2.97 \times 10^{-105}$ \\
& nonRE oxide -- nonRE nonoxide  & 748 & 785  & 269691.0 & $5.81 \times 10^{-3}$ \\

\hline
\end{tabular}
\end{table}

\begin{equation}
\frac{\sigma_f}{f}
=
\sqrt{
\left(
\frac{\sigma_{\Theta}}
{|\Theta|}
\right)^2
+
\left(
\frac{\sigma_{T_\mathrm{N}}}
{T_\mathrm{N}}
\right)^2
}.
\label{eq:errorprop}
\end{equation}

\begin{table}[H]
\centering
\caption{Prediction performance of the binary subdivision models.}
\label{tab:binary_models}
\small
\begin{tabular}{llccc}
\hline
Property & Metric & CV & OOF & Test\\
\hline

\multirow{3}{*}{$T_\mathrm{N}$ (Oxides)}
&MAE&55.21&55.22&54.27\\
&RMSE&101.82&108.41&95.05\\
&$R^2$&0.714&0.689&0.760\\
\hline

\multirow{3}{*}{$T_\mathrm{N}$ (Non-oxides)}
&MAE&40.83&40.82&38.13\\
&RMSE&86.22&86.87&109.34\\
&$R^2$&0.622&0.635&0.528\\
\hline

\multirow{3}{*}{$\Theta$ (RE)}
&MAE&42.52&42.51&34.71\\
&RMSE&89.42&90.77&83.98\\
&$R^2$&0.599&0.601&0.715\\
\hline

\multirow{3}{*}{$\Theta$ (Non-RE)}
&MAE&99.57&99.57&83.87\\
&RMSE&167.69&170.77&165.83\\
&$R^2$&0.539&0.539&0.679\\
\hline
\end{tabular}
\end{table}

\begin{figure}[H]
    \centering
    \includegraphics[width=0.9\linewidth]{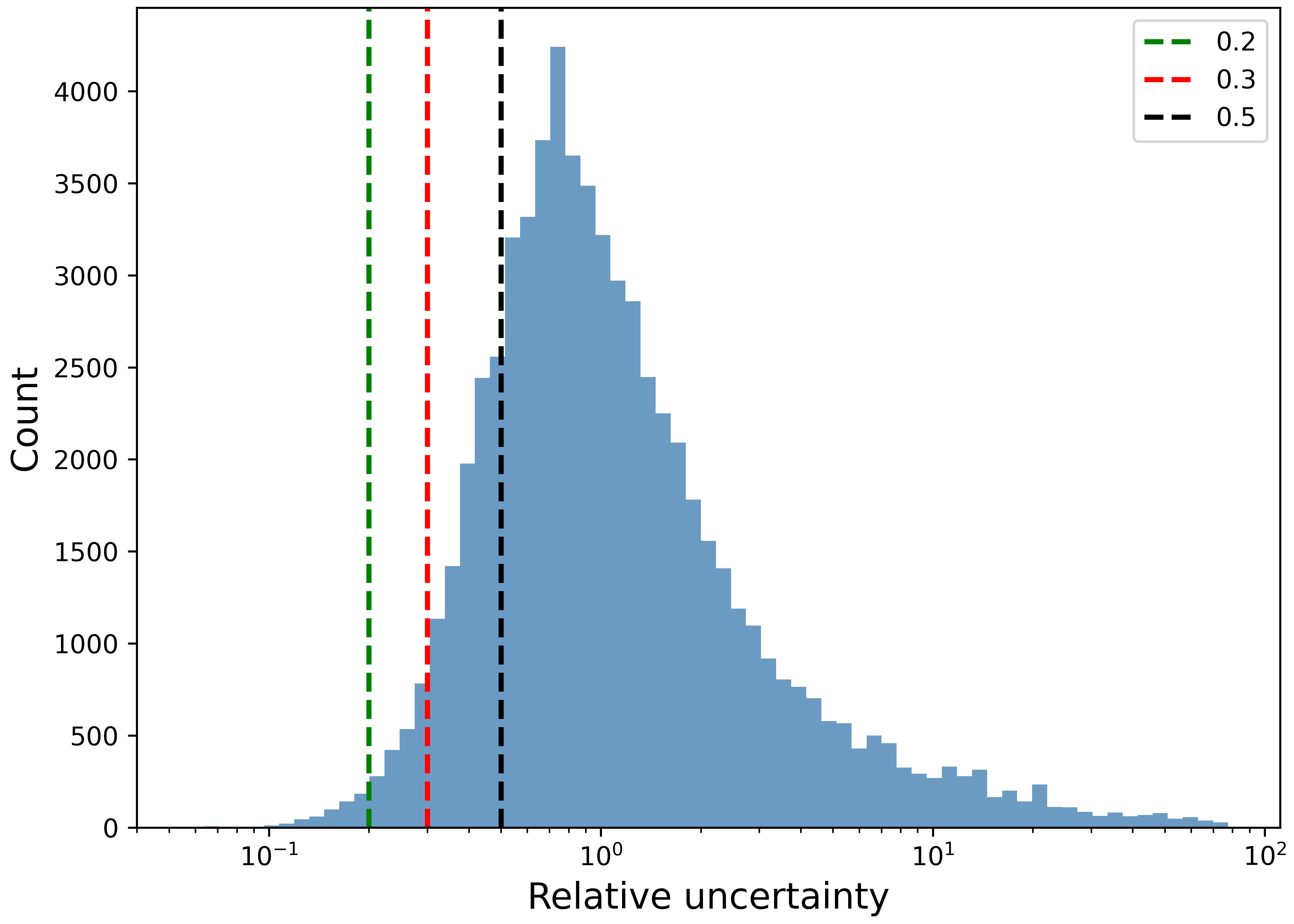}
    \caption{Distribution of the relative prediction uncertainty estimated from the ensemble variance for all screened compounds.}
    \label{fig:chemical_classes}
\end{figure}

\begin{figure}[htbp]
\centering
\includegraphics[width=0.85\linewidth]{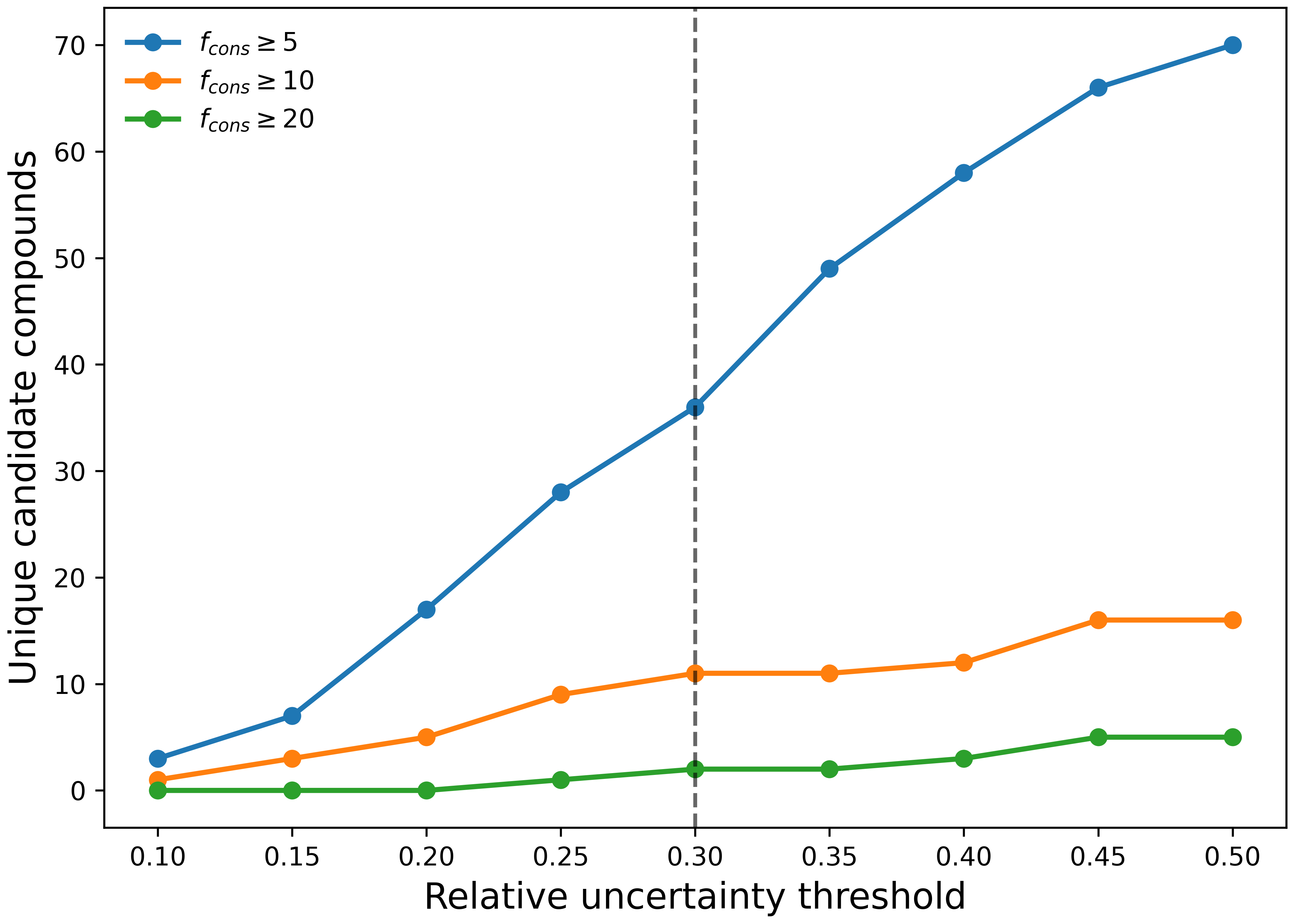}
\caption{\textcolor{blue}{Sensitivity of the candidate screening results to the relative-uncertainty threshold. The number of unique candidate compounds satisfying \(f_{\mathrm{cons}}\geq 5\) is shown as a function of the relative-uncertainty threshold. Candidate formulas were normalized and duplicates removed before counting. The threshold of 0.3 was selected as a practical compromise between uncertainty control and screening coverage.}}
\label{fig:TN_interpolation_extrapolation}
\end{figure}

\begin{figure}[H]
    \centering
    \includegraphics[width=0.9\linewidth]{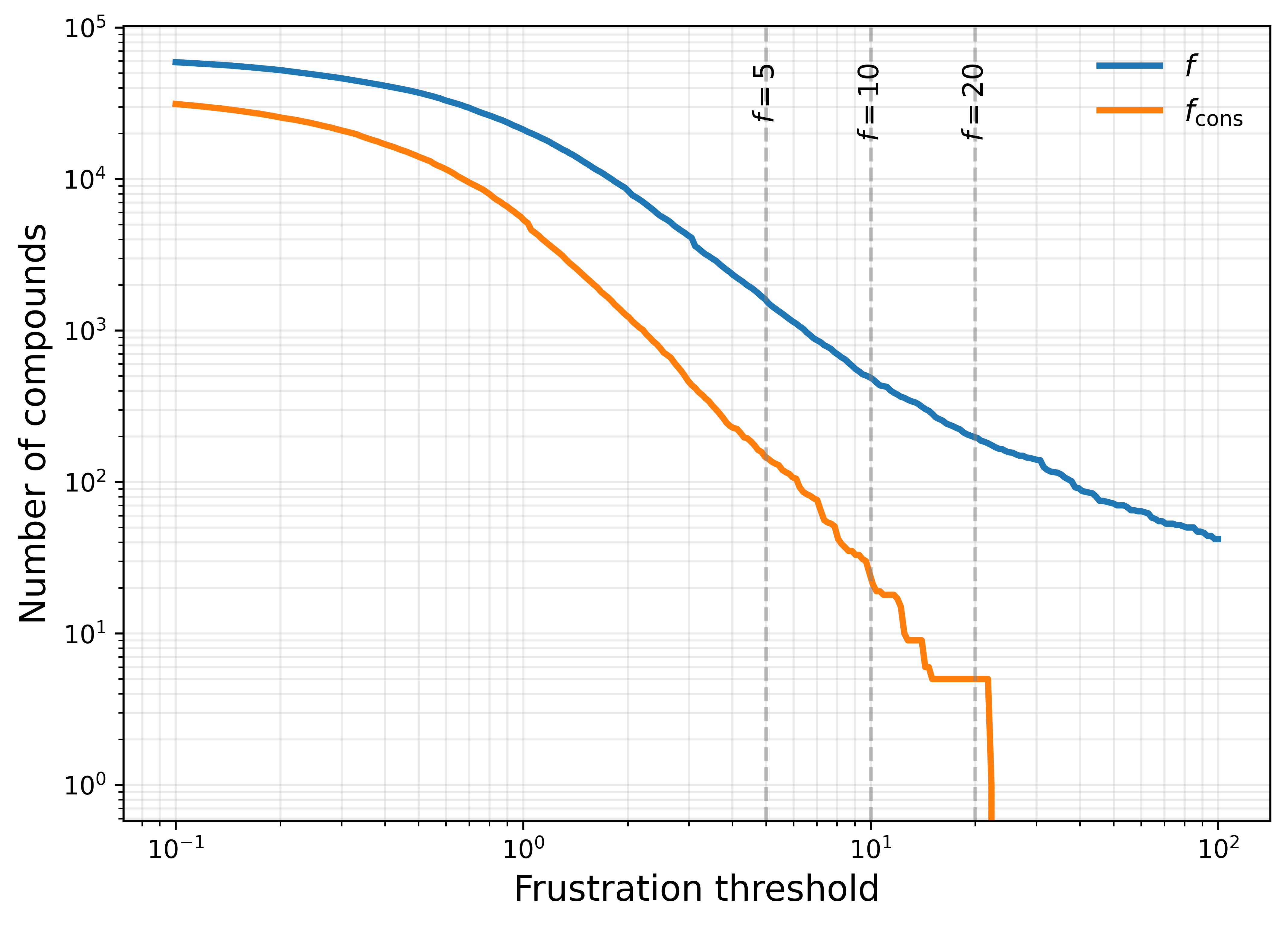}
    \caption{Number of screened compounds exceeding different frustration thresholds calculated using the conventional frustration index ($f$) and the uncertainty-aware conservative frustration index ($f_{\mathrm{cons}}$).}
    \label{fig:chemical_classes}
\end{figure}

\begin{figure}[H]
\label{fig:first}
    \centering
    \begin{minipage}{0.8\textwidth}
        \centering
        \includegraphics[width=\linewidth]{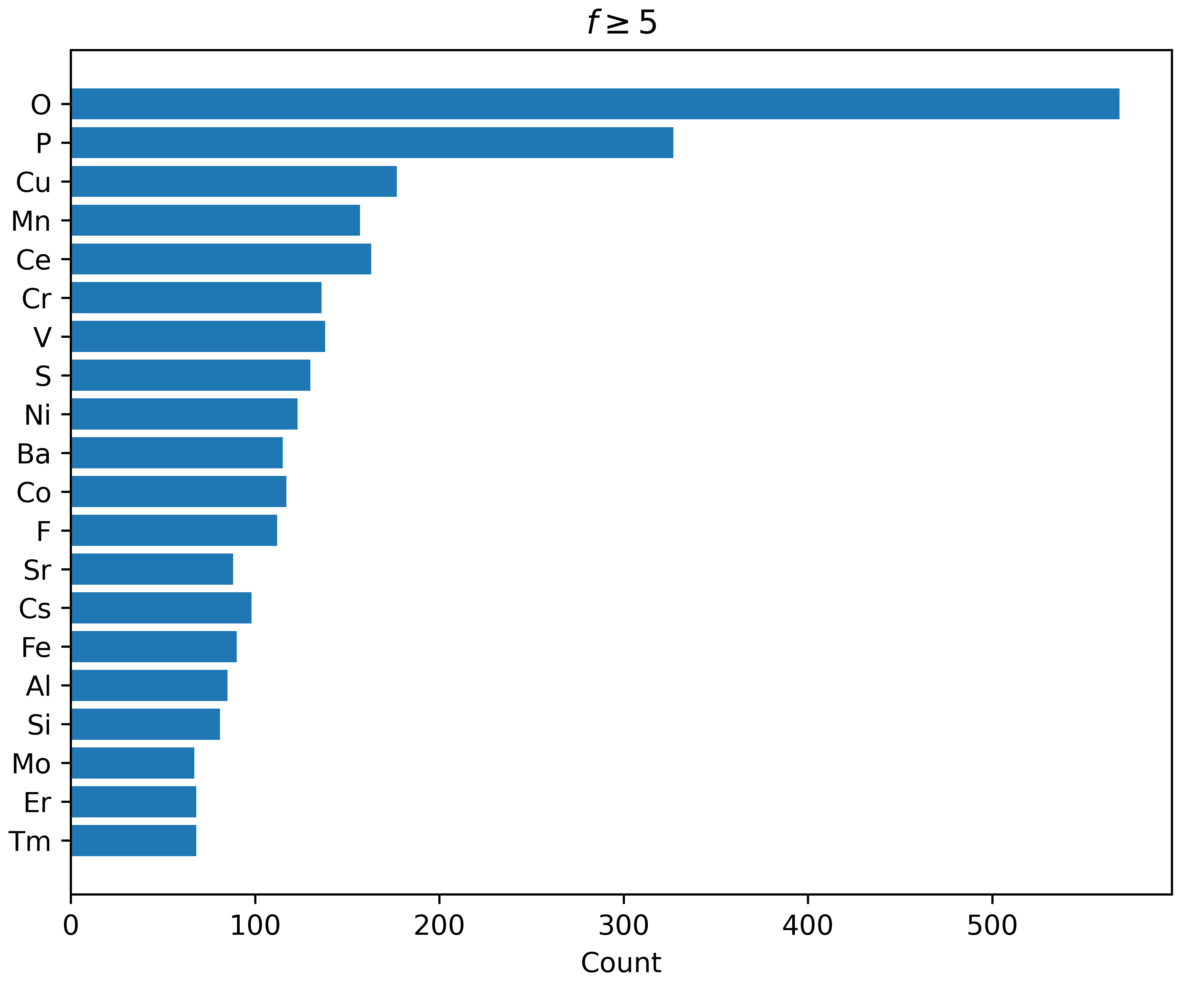} \\
        (a)
           \end{minipage} 
    
    \begin{minipage}{0.8\textwidth}
        \centering
        \includegraphics[width=\linewidth]{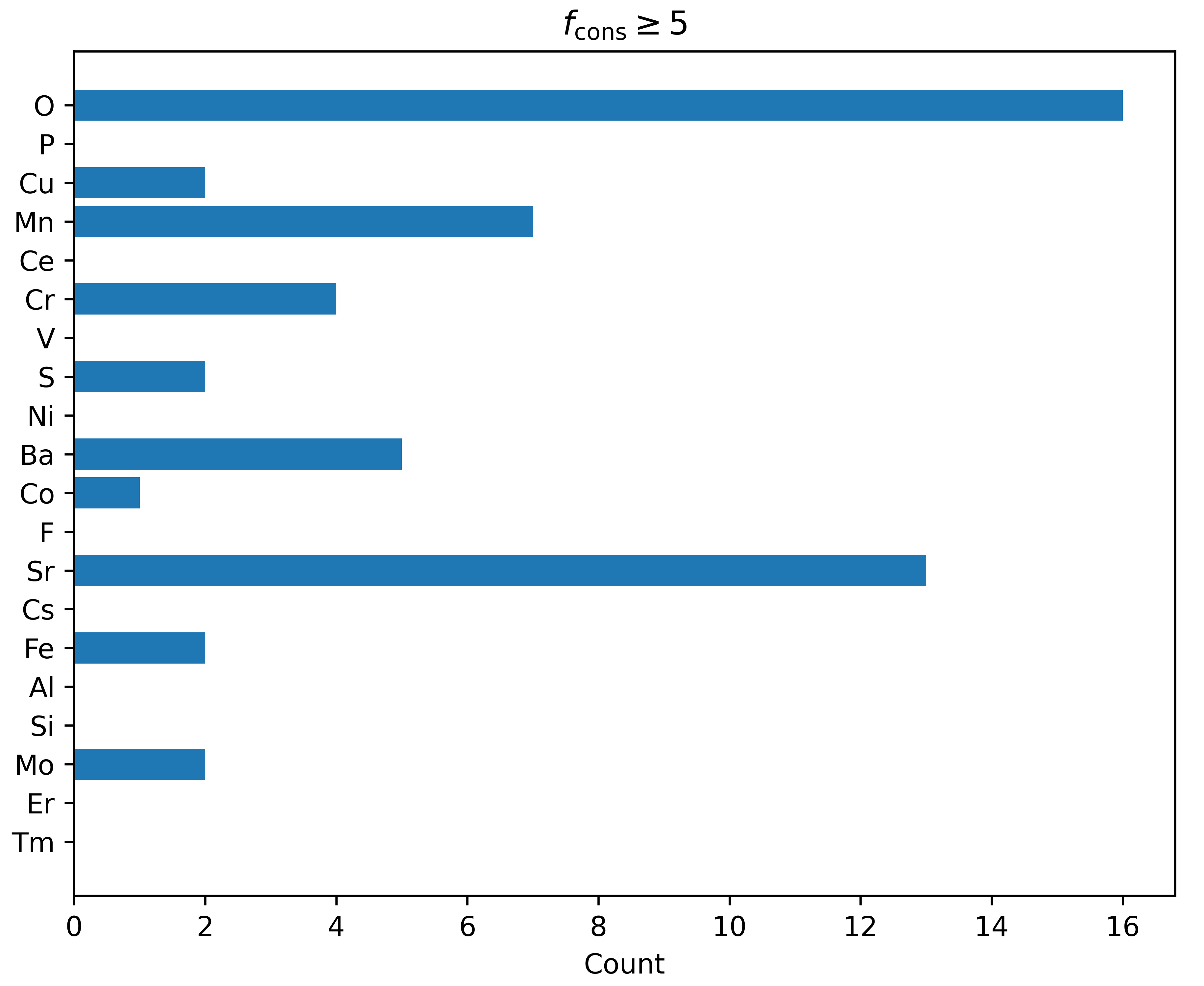} \\
        (b)
          \end{minipage}
    \caption{Comparison of elemental distributions for compounds satisfying (a) the conventional frustration criterion ($f \geq 5$) and (b) the uncertainty-aware conservative criterion ($f_{\mathrm{cons}} \geq 5$).}
\end{figure}

\begin{longtable}{lrrrrr}
\caption{Final high-confidence frustrated magnetic candidates identified after excluding all compositions present in the training datasets.}
\label{tab:final_candidates}\\

\toprule
Formula & $T_N$ (K) & $\Theta$ (K) & $f$ & $f_{\mathrm{cons}}$ & Relative uncertainty \\
\midrule
\endfirsthead

\multicolumn{6}{c}{{\bfseries Table \thetable\ (continued from previous page)}}\\
\toprule
Formula & $T_N$ (K) & $\Theta$ (K) & $f$ & $f_{\mathrm{cons}}$ & Relative uncertainty \\
\midrule
\endhead

\midrule
\multicolumn{6}{r}{\emph{Continued on next page}}\\
\endfoot

\bottomrule
\endlastfoot

Sr$_{2}$NbCoO$_{6}$ & 19.3 & 602.9 & 31.30 & 21.81 & 0.260 \\
ZnCr$_{8}$Cu$_{3}$Se$_{16}$ & 21.8 & 377.7 & 17.36 & 14.74 & 0.132 \\
Sr$_{3}$CaTi$_{2}$Mn$_{2}$O$_{12}$ & 42.0 & -688.8 & 16.39 & 14.05 & 0.152 \\
SrCaTiMnO$_{6}$ & 34.8 & -568.0 & 16.31 & 12.38 & 0.227 \\
Sr$_{7}$CaTi$_{5}$Mn$_{3}$O$_{24}$ & 40.9 & -593.9 & 14.53 & 12.11 & 0.130 \\
Sr$_{5}$Ca$_{3}$Ti$_{5}$Mn$_{3}$O$_{24}$ & 35.1 & -552.9 & 15.73 & 12.02 & 0.202 \\
Sr$_{2}$FeMoO$_{6}$ & 34.5 & 473.5 & 13.74 & 9.86 & 0.267 \\
Ba(Cr$_{5}$S$_{8}$)$_{2}$ & 55.8 & -636.7 & 11.41 & 9.68 & 0.121 \\
Sr$_{7}$CaTi$_{3}$Mn$_{5}$O$_{24}$ & 48.6 & -584.5 & 12.02 & 8.56 & 0.246 \\
NaSr$_{9}$Fe$_{5}$(MoO$_{6}$)$_{5}$ & 37.2 & 391.5 & 10.53 & 7.99 & 0.226 \\
ZnCr$_{4}$CuSe$_{8}$ & 25.9 & 258.4 & 9.97 & 7.91 & 0.166 \\
BaCr$_{4}$S$_{7}$ & 36.7 & -386.2 & 10.51 & 7.54 & 0.249 \\
Ba$_{2}$LuIrO$_{6}$ & 35.8 & -379.4 & 10.61 & 7.20 & 0.298 \\
Sr$_{2}$YOsO$_{6}$ & 46.9 & -416.8 & 8.88 & 6.75 & 0.200 \\
Sr$_{2}$ScOsO$_{6}$ & 45.5 & -385.5 & 8.47 & 6.49 & 0.198 \\
Ba$_{2}$YIrO$_{6}$ & 47.5 & -401.2 & 8.45 & 6.39 & 0.200 \\
Ba$_{2}$ScIrO$_{6}$ & 52.6 & -415.4 & 7.90 & 6.34 & 0.154 \\
Sr$_{7}$CaTi$_{7}$MnO$_{24}$ & 45.5 & -385.9 & 8.47 & 6.07 & 0.252 \\
Sr$_{3}$CaTiMn$_{3}$O$_{12}$ & 51.9 & -452.4 & 8.71 & 5.83 & 0.276 \\
Sr$_{2}$LuIrO$_{6}$ & 41.3 & -328.5 & 7.95 & 5.52 & 0.262 \\

\end{longtable}